\documentclass[10pt,aps,prl,superscriptaddress,twocolumn,showpacs,floatfix,notitlepage]{revtex4-1}
\usepackage{amsmath}
\usepackage{amsfonts}
\usepackage{amssymb}
\usepackage{hyperref}
\usepackage[shortcuts]{extdash}
\usepackage{graphicx}
\usepackage{color}
\usepackage{xcolor}
\usepackage[mathscr]{euscript}
\usepackage{natbib}
\usepackage{bbm}
\usepackage{rotating}

\newcommand{\bra}[1]{\left\langle #1\right|}
\newcommand{\ket}[1]{\left| #1\right\rangle}

\newcommand{\ud}[1]{{#1^{\dagger}}}
\newcommand{\av}[1]{{\langle #1 \rangle}}
\newcommand{\mcdots}{\mathinner{\cdotp\!\cdotp\!\cdotp}}

\hypersetup{colorlinks=true,linkcolor=magenta
  ,citecolor=magenta, urlcolor=magenta}

\begin{document} 
\flushbottom 

\title{Multiphoton Emission}

\author{G.~D\'{i}az Camacho}
\thanks{These two authors contributed equally}
\affiliation{Departamento de F\'isica Te\'orica de la Materia
Condensada \& IFIMAC, Universidad Aut\'onoma de Madrid, 28049 Madrid,
Spain} 

\author{E.~Zubizarreta Casalengua}
\thanks{These two authors contributed equally}
\affiliation{Faculty of Science and
  Engineering, University of Wolverhampton, Wulfruna St, Wolverhampton
  WV1 1LY, UK}
\affiliation{Departamento de F\'isica Te\'orica de la Materia
Condensada \& IFIMAC, Universidad Aut\'onoma de Madrid, 28049 Madrid,
Spain} 

\author{J.~C.~{L\'{o}pez~Carre\~{n}o}}
\affiliation{Institute of Theoretical Physics, University of Warsaw, ul. Pasteura 5, 02-093, Warsaw, Poland
}
\author{S.~Khalid}
\affiliation{Faculty of Science and
  Engineering, University of Wolverhampton, Wulfruna St, Wolverhampton
  WV1 1LY, UK}

\author{C.~Tejedor}
\affiliation{Departamento de F\'isica Te\'orica de la Materia
Condensada \& IFIMAC, Universidad Aut\'onoma de Madrid, 28049 Madrid,
Spain} 

\author{E.~del~Valle}
\affiliation{Departamento de F\'isica Te\'orica de la Materia
Condensada \& IFIMAC, Universidad Aut\'onoma de Madrid, 28049 Madrid,
Spain} 
\affiliation{Institute for Advanced Study, Technical University of Munich, Lichtenbergstrasse 2a, D-85748 Garching, Germany.}
\affiliation{Faculty of Science and
  Engineering, University of Wolverhampton, Wulfruna St, Wolverhampton
  WV1 1LY, UK}

\author{F.~P.~Laussy}
\affiliation{Faculty of Science and 
  Engineering, University of Wolverhampton, Wulfruna St, Wolverhampton
  WV1 1LY, UK}
\affiliation{Russian Quantum Center, Novaya 100, 143025 Skolkovo,
  Moscow Region, Russia}
\email{fabrice.laussy@gmail.com}

\date{\today}

\begin{abstract}
  We describe the emission, detection and structure of multiphoton
  states of light. By including frequency filtering, we describe at a
  fundamental level the physical detection of the quantum state. The
  case of spontaneous emission of Fock states is treated fully and
  analytically.  We stress this picture by contrasting it to the
  numerical simulation of two-photon bundles emitted from a two-level
  system in a cavity. We show that dynamical factors exist that allow
  for a more robust multiphoton emission than spontaneous emission.
  We also describe how thermal light relates to multiphoton~states.
\end{abstract}

\maketitle

%\cite{hofheinz09a}
%

Multiphoton (or multiquanta) physics is rapidly emerging in various
areas of quantum technology~\cite{deleglise08a, hofheinz08a,
  sayrin11a, chu18a, sanchezmunoz20a, vinasbostrom20a, zhong20a,
  uria20a}.  In this Letter, we provide a detailed description of
basic and fundamental aspects of multiphoton emission and
detection. Namely, we give exact closed-form expressions for the full
photon-counting probabilities of the frequency-filtered Spontaneous
Emission (SE) of Fock states, from which we derive their complete
temporal structure and thus all statistical observables of possible
interest. We discuss particular cases for illustration and then
contrast SE to Continuous Wave (CW) emission and thermal equilibrium.

To give the most general description, we compute the
probability~$p(n,T;t)$ of detecting $n$~photons in the time window
between $t$ and~$t+T$. For a single-mode source with bosonic
annihilation operator~$a$ and emission rate~$\gamma_a$, this is given
by Mandel's formula~\cite{mandel59a,kelley64a}
%
% \begin{equation}
%   \label{eq:Fri14May173224CEST2021}
%   p(n,T;t) = \frac{1}{n!}  \left\langle : \Omega^n \exp (-\Omega) :
% \right\rangle
% \end{equation}
% %
$p(n,T;t) = \frac{1}{n!}  \left\langle : \Omega^n \exp (-\Omega) :
\right\rangle$ with $:\kern.1cm:$ denoting normal ordering and
$\Omega$ the time-integrated intensity operator
$\Omega(t,T)\equiv\xi\gamma_a \int_{t}^{t+T} \left(\ud{a} a \right)
(t') \, dt'$, $\xi$ quantifying detection efficiency (1 for a perfect
detector).  This is more general than the popular Glauber correlation
functions~$g^{(n)}$~\cite{glauber63a}, that can be derived from
it. This is apparent in the expanded form for
$p(n,T;t) = \frac{(-1)^n}{n!} \sum_{k=0}^{\infty}
\frac{(-\xi\gamma)^{n+k} (n+k)!}{k!}
\int_{t}^{t+T}\int_{t}^{t3}\int_{t}^{t2}\langle \ud{a} (t_1) \dots
\ud{a} (t_{n+k})\break a (t_{n+k}) \dots a(t_1) \rangle dt_1 dt_2\dots
dt_{n+k}$. We first compute it for the simplest and most fundamental
type of multiphoton emission, namely, that of a group, or
``bundle''~\cite{sanchezmunoz14a, dong19a, bin20a, bin21a, ma21a,
  deng21a, arXiv_cosacchi21a}, of $N$~photons, emitted spontaneously
by a source, e.g., a passive cavity of natural frequency $\omega_a$ in
the state~$\rho(0) = \ket{N}\bra{N}$ at~$t=0$, that freely radiates
its photons at the rate~$\gamma_a$ without any type of stimulation or
other dynamical factor. Under these conditions, the state evolves
according to the Lindblad master equation (we set~$\hbar=1$)
$\dot{\rho} = -i \, \omega_a [ \ud{a} a, \rho] + \frac{\gamma_a}{2}(2
a \rho \ud{a} - \ud{a} a \rho - \rho \ud{a} a)$ for which one can
find, % the multi-time correlators
% $\av{{:}(\prod_{k=1}^{n} a(t_k))^\dagger\prod_{k=1}^{n} a(t_k){:}}$,
% e.g., for $m=1$, one has
% $\av{\ud{a}(t'_1) a(t_1)} =\av{\ud{a}a}(0) \, e^{-(\gamma_a/2 - i
%   \omega_a) t'_1} e^{-(\gamma_a/2 + i \omega_a) t_1}$ regardless of
% the time order. For SE, this can,
through a tedious but systematic procedure~\cite{multisup}, the
multi-time correlators
$\av{\ud{a}(t'_1)\dots\ud{a}(t'_m) a(t_m)\dots(t_1)} = \
\frac{N!}{(N-m)!}\big(\prod_{\kappa=1}^m e^{-(\gamma_a/2 - i \omega_a)
  t'_{\kappa}} \big) \big( \prod_{k=1}^m e^{-(\gamma_a/2 + i \omega_a)
  t_k} \big)$.~One can then compute the quantum-average for
any power of $\Omega$ by splitting its time-dynamics from the operator
$a^\dagger a$ in a scalar function~$\mathscr{T}$ as $\Omega(t, T)
= % \xi \gamma_a \int_0^T (\ud{a} a) (t) dt =
\mathscr{T}(T)\ud{a}(t)a(t)$.
%
%%
%\begin{equation}
%\Omega(0, T) = % \xi \gamma_a \int_0^T (\ud{a} a) (t) dt = 
%\mathscr{T}(T)\ud{a}(0)a(0)
%\end{equation}
%%
%
For SE, with
$\Omega(0,T)= \gamma_a\xi \left( \int_0^T e^{-\gamma_a t} dt \right)
\ud{a}(0)a(0)$, it gives:
\begin{equation}
  \label{eq:Tue18May121411CEST2021}
    \mathscr{T}(T)=\xi(1-e^{-\gamma_a T})\,.
\end{equation}
With time and quantum-averages now separated, it is straightforward to
compute
$\av{: \Omega^k :} \ = \mathscr{T}^k \av{a^{\dagger k} a^k}(0) =\xi^k
\left(1- e^{-\gamma_a T}\right)^k \frac{N!}{(N-k)!}$ for all
$1\le k\le N$.  Substituting in Mandel's formula (now keeping track in
the notation of the initial state~$N$ instead of the initial
time~$t$),
$p(n,T;N) = \sum_{k=n}^{\infty} {(-1)^{n+k}
  \av{:\Omega^k:}}/(n!(k-n)!)$, this simplifies to the physically
transparent form:
\begin{equation}
  \label{eq:Thu3Dec150332CET2020}
  p(n,T;N) = {N\choose n}\mathscr{T}(T)^n(1-\mathscr{T}(T))^{N-n}\,,
\end{equation}
that is, a binomial distribution arising from independent Bernoulli
trials for the detection of a single photon in the
time~$T$. Variations of this result have been known for a long
time~\cite{scully69a, arnoldus84a} and, with hindsight,
Eqs.~(\ref{eq:Tue18May121411CEST2021})
and~(\ref{eq:Thu3Dec150332CET2020}) could  have been postulated
based on physical arguments.
Equation~(\ref{eq:Tue18May121411CEST2021}) is indeed the probability
to detect a single photon in SE in the time window~$[0,T]$ and thus
corresponds to the quantum efficiency.

\begin{figure}[t!]
  \includegraphics[width=\linewidth]{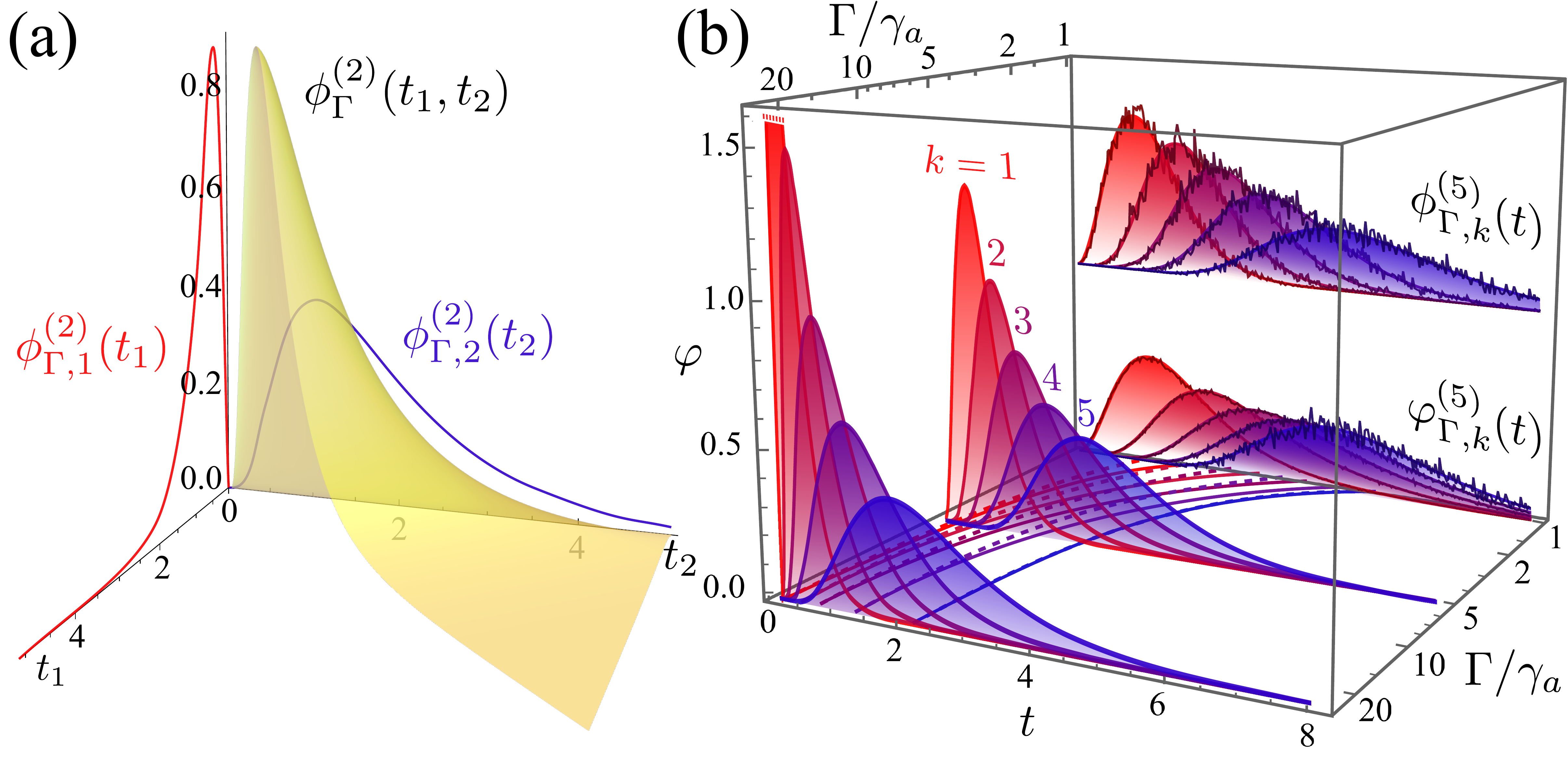}
  \caption{Filtered $N$-photon spontaneous emission. (a) For $N=2$
    photons, one can visualize the full pdf
    $\phi^{(2)}_\Gamma(t_1,t_2)=2\gamma_a^2(\Gamma/\Gamma_-)^4g'(t_1)g'(t_2)\mathbf{1}_{[t_1,\infty[}(t_2)$
    in addition to its marginals~$\phi_{\Gamma,k}^{(2)}(t_k)$ which
    are its projections on the single-photon subspaces. (b) Marginals
    for~$N=5$ for various filter widths~$\Gamma$, conditioned ($\phi$,
    up) or not ($\varphi$, down) to full-bundle detection. The
    case~$\Gamma/\gamma_a=1$ has superimposed a numerical
    simulation. On the floor, the mean
    times~(\ref{eq:Mon16Aug170254CEST2021}) showing dilation of the
    bundles under filtering (solid for~$\varphi$, dashed for~$\phi$).}
  \label{fig:Thu27May142405CEST2021}
\end{figure}

The value of the above derivation is, besides a rigorous and direct
justification of the result, its extension to the case of filtered
emission. At a fundamental level, frequency filtering describes
detection~\cite{eberly77a, delvalle12a}, since any physical detector
has a finite temporal resolution and, consequently, a frequency
bandwidth~$\Gamma$, which restricts the quantum attributes of the
measured system. Ultimately, any quantum system has to be observed
and, at this point, instead of being a mere technical final step,
detection typically brings all the counter-intuitive and disruptive
nature of the theory.  Including these fundamental constrains is
therefore essential for a physical theory of multiphoton emission.  To
do that, the previous derivation can be repeated but now involving
filtered-field operators
$\varsigma (t) \equiv \int_{-\infty}^{\infty} \mathcal{F}(t - t_1)
a(t_1) \, dt_1$
% %
% \begin{equation}
%   \label{eq:Fri4Dec171242CET2020}
%   \varsigma (t) \equiv \int_{-\infty}^{\infty} \mathcal{F}(t - t_1) a(t_1) \, dt_1 \,,
% \end{equation}
%
whereby a filter~$\mathcal{F}$ is applied to the source's field.  For
SE, the time dynamics can similarly be separated from the quantum
operator $\varsigma (t)=\Xi(t) a(0)$ that now combines both the
dynamics of filtering and SE in
$\Xi\equiv\int_{-\infty}^{+\infty}\mathcal{F}(t-t_1)e^{-(\frac{\gamma_a}2+i\omega_a)
  t_1}\theta(t_1)dt_1$ ($\theta$ the Heavyside function). In this
way, the filtered time-integrated intensity operator from $[0,T]$
reads
$\Omega_\Gamma(T) = \gamma_a\xi \int_{0}^{T} \ud{\varsigma} \varsigma
(t) \, dt = \mathscr{T}_\Gamma(T)(\ud{a} a)(0)$. This provides the
single-photon-detection probability (quantum efficiency) for a
filtered-field
$ \mathscr{T}_\Gamma(T)=\gamma_a\xi \int_{0}^{T}
\Xi_{\mathcal{F}}(t)^* \, \Xi_{\mathcal{F}}(t) \, dt$.
%
% \begin{equation}
%   \label{eq:Wed19May102632CEST2021}
%   \mathscr{T}_\Gamma(T)=\gamma_a \int_{0}^{T} \Xi_{\mathcal{F}}(t)^* \, \Xi_{\mathcal{F}}(t) \, dt\,.
% \end{equation}
%
For interference (Lorentzian) filters,
$\mathcal{F}(t) = \frac{\Gamma}{2} e^{-(\Gamma/2 + i \, \omega_a) t}
\, \theta(t)$, and for SE,
$\Xi(t)=(\Gamma/\Gamma_-)({e^{-(\gamma_a/2 + i \, \omega_a) t} -
  e^{-\Gamma t/2}})\theta(t)$, in which case,
defining $\Gamma_{\pm} \equiv \Gamma \pm \gamma_a$, the quantum
efficiency to use in Eq.~(\ref{eq:Thu3Dec150332CET2020}) for a
physical detector, or to describe the filtering of light, is:
\begin{multline}
  \label{eq:Fri4Dec172556CET2020}
\mathscr{T}_\Gamma(T)=
  \frac{\Gamma}{\Gamma_+} - \frac{\Gamma^2 \, e^{- \gamma_a T} + \Gamma\gamma_a \, e^{- \Gamma T}}{\Gamma_-^2}+ \frac{4 \Gamma^2 \gamma_a e^{- \Gamma_+ T/2} }{\Gamma_-^2 \Gamma_+}\,.
\end{multline}
It generalizes Eq.~(\ref{eq:Tue18May121411CEST2021}), which it
recovers in the limit~$\Gamma\to\infty$ since, due to Heisenberg
uncertainty, one must detect at all the frequencies and loose this
information, to know precisely when the photons are being detected. On
the other hand, the~$T\to\infty$ limit gives
$P(k,N)\equiv\lim_{T\to\infty}p(k,T;N)={N\choose
  k}{\gamma_a^{N-k}\Gamma^k\over\Gamma_+^N}$ as the probability to
detect~$k$ out of~$N$ emitted photons. This is less than~1 even for a
perfect detector~($\xi=1$), due to its finite bandwidth, showing again
the fundamental interplay of time and frequency in the detection. We
can now describe completely the SE of~$N$ photons. This comes in the
form of the joint probability distribution function
(pdf)~$\phi^{(N)}(\{t_k\})$ for the~$k$th photon to be detected
\emph{at} time~$t_k$. We derive this quantity from its link to the
photon counting probabillity by defining $p(t_1,\dots,t_N)$ the
probability to detect the~$k$th photon \emph{up to} time~$t_k$ as
$p(\{t_k\})=\int_0^{t_1}\dots\int_0^{t_N}\phi^{(N)}(\{\tau_k\})\prod_{i=1}^Nd\tau_i$. Since
this is also related to the combinatorics of single-photon detection
probabilities, namely,
$p(\{t_k\})=N!\mathscr{T}(t_1)\prod_{k=1}^{N-1}\big(\mathscr{T}(t_{k+1})-\mathscr{T}(t_{k})\big)$~\cite{multisup},
one can finally
obtain~$\phi(t_1,\dots,t_N)={\partial^N\over\partial t_1\cdots\partial
  t_N}p(t_1,\dots,t_N)$ from the fundamental theorem of calculus.  We
give directly the filtered emission result, from which unfiltered
emission arises as the limiting case~$\Gamma\to\infty$ (it is
explicited in~\cite{multisup}):
\begin{multline}
  \label{eq:Mon7Jun131211CEST2021}
  \phi_\Gamma^{(N)}(t_1,\dots,t_N)=N! \gamma_a^N\times{}\\
  \left(\frac{\Gamma}{\Gamma_-}\right)^{2N }\prod_{i=1}^{N}
  (e^{-\Gamma t_i/2}-e^{-\gamma_a t_i/2})^2{\mathbbm{1}}_{[t_{i-1},
    t_{i+1}[}(t_{i})
\end{multline}
where~$\mathbbm{1}_T(t)$ is the indicator function which is~$1$
if~$t\in T$ and~$0$ otherwise.  This joint probability distribution
provides the exact and complete temporal structure of multiphoton SE,
and is one of the main results of this work (the~$N=2$ case is shown
in Fig.~\ref{fig:Thu27May142405CEST2021}(a)). From it, one can compute
all the statistical quantities regarding filtered multiphoton SE. As
an illustration, we derive the marginal distributions for the emission
of each photon in isolation
$\phi_{\Gamma,k}^{(N)}(t_k)\equiv\idotsint\phi(t_1,\dots,t_N)\prod_{j\neq
  k} dt_j$ and, from there, emission time
averages~$\langle t_k^{(N)}\rangle\equiv\int_0^\infty
t_k\phi_{\Gamma,k}^{(N)}(t_k)dt_k$ (others like variances, etc., are
illustrated in the Supplementary Material~\cite{multisup}).  Although
less fundamental and general, one-photon marginals are more familiar,
accessible and practical. We find
$\phi^{(N)}_{\Gamma,k}(t_k)\equiv-{\left({\Gamma\over\Gamma_-}\right)^{2N}}\gamma_a^Nk{N\choose
  k}g(t_k)^{N-k}(g(0)-g(t_k))^{k-1}g'(t_k)$~where
%~\cite{multisup}:
%
% \begin{multline}
%   \label{eq:Thu17Jun114723CEST2021}
%   \phi^{(N)}_{\Gamma,k}(t_k)=-{\left({\Gamma\over\Gamma_-}\right)^{2N}}\gamma_a^N\times\\k{N\choose k}g(t_k)^{N-k}(g(t_0)-g(t_k))^{k-1}g'(t_k)
% \end{multline}
%
we defined
$g(t)\equiv {e^{-\gamma_a t}\over\gamma_a}+{e^{-\Gamma
    t}\over\Gamma}-4{e^{-\Gamma_+t/2}\over\Gamma_+}$ so that
$g'(t)=-(e^{-\Gamma t/2}-e^{-\gamma_a t/2})^2$---the term that enters
in Eq.~(\ref{eq:Mon7Jun131211CEST2021})---makes the integration of the
marginal tractable in the form of nested integrals~\cite{multisup}
$\int_{t_0}^{t_k}\mcdots\int_{t_{k-2}}^{t_k}\int_{t_k}^\infty\mcdots\int_{t_{N-1}}^\infty\prod_{i=1}^N
g'(t_i)dt_1\mcdots dt_{k-1}dt_{k+1}\mcdots d_{t_N}$.  The
distributions~$\phi^{(N)}_{\Gamma,k}(t_k)$ are for the $k$th photon
from a fully-detected $N$-photon bundle. To take into account that
filtering removes some photons, one needs to turn to the conditional
probabilities~$\varphi_k^{(N)}(t)$ of detecting the~$k$th photon
regardless of how many photons have been detected. This is obtained as
the probability to detect the photon in the~$k$th position in a
$N$-photon bundle of which~$n$ photons have been detected, the
probability of which is Eq.~(\ref{eq:Thu3Dec150332CET2020}), so that,
from the law of total probability~\cite{multisup},
$
\varphi_{n}^{(N)}(t_n)\equiv\sum_{k=n}^{N}\frac{P(k,N)}{P(k,k)}\phi_{\Gamma,n}^{(k)}(t_n)$. 
% %
% \begin{equation}
%   \label{eq:Thu10Jun125725CEST2021}
%   \varphi_{n}^{(N)}(t_n)=\sum_{k=n}^{N}\frac{P(k,N)}{P(k,k)}\phi_{\Gamma,n}^{(k)}(t_n)\,,
% \end{equation}
%
Here, $\varphi_n^{(N)}$ is normalized to the fraction of
$n$-photon-bundles detected in the SE of $N$ photons,
$
\mathcal{N}(n,N)\equiv\sum_{k=n}^NP(k,N)={\Gamma^n\gamma_a^{N-n}\over\Gamma_+^N}{N\choose
  n}{}_2F_1(1,n-N,n+1,{-\Gamma\over\gamma_a})$ where ${}_2F_1$ is the
Hypergeometric function.  Qualitatively, the detected photons are
delayed and spread, with earlier photons being more affected. A
representative case is shown for~$\varphi$ in
Fig.~\ref{fig:Thu27May142405CEST2021}(b) for~$N=5$ and filter
widths~$\Gamma/\gamma_a=1, 5$ and~$25$, the latter case of which
starts to be close to unfiltered emission. Comparing to a
full-detection case~$\phi$, shown in the upper part, one sees that
incomplete bundles further spread out photons and pile them up towards
later times. Note that the 5th photon distribution is the same in both
cases since these are detected only when the bundle remains unbroken.
From all the possible statistical observables that one can derive from
these results (as illustrated in the Supplementary
Material~\cite{multisup}), it should be enough here to contemplate the
mean time of detection for the~$k$th photon of a $N$-photon bundle,
which, although it could seem a simple quantity, actually requires an
11-indices summation
$\sum_{\{k_i\}}\equiv\sum_{k_1+k_2+k_3\atop=N-k}\sum_{k_1+\cdots+k_9\atop=k-1}\sum_{k_{10}+k_{11}\atop=2}$:
\begin{multline}
  \label{eq:Mon16Aug170254CEST2021}
  \langle t_{k}^{(N)} \rangle = 2 N!\left({\gamma_a\Gamma\Gamma_+\over\Gamma_-^2}\right)^N\sum_{\{k_i\}}\prod_{j=1}^{11}{1\over k_j!}\times{}\\ \frac{(-1)^{k_3 + k_6 + k_7 + k_8 + k_{11}}}{\gamma_a^{\Sigma_1} \Gamma^{\Sigma_2} \left( \frac{\Gamma_+}{4} \right)^{\Sigma_3} \big( \gamma_a\Sigma_4 +\Gamma\Sigma_5 + \frac{\Gamma_+}{2} (k_3+k_9)\big)^2 }
\end{multline}
where~$\Sigma_1\equiv k_1+k_4+k_7$, $\Sigma_2\equiv{k_2+k_5+k_8}$,
$\Sigma_3\equiv k_3 + k_6 + k_9$,
$\Sigma_4\equiv k_1 + k_7+{k_{11}\over2}$ and
$\Sigma_5\equiv k_2 +k_8 +{k_{10}\over2}$.
This generalizes the unfiltered-photons result
$\lim_{\Gamma\to\infty}\langle t_k^{(N)}\rangle =
(H_N-H_{N-k})/\gamma_a$ (with~$H_N$ the $N$th harmonic number and
assuming~$H_0\equiv 0$) and explains the observed photon-bundle time
length (the time between the first and last detected
photons)~$\langle\tau\rangle_N\equiv\langle t_N^{(N)}\rangle-\langle
t_1^{(N)}\rangle$ that goes to $H_{N-1}/\gamma_a$ in this limit, as
was observed in Refs.~\cite{sanchezmunoz14a, arXiv_cosacchi21a}, as
well as the~$k$th photon lifetime measured as
$\langle t_k^{(N)}\rangle-\langle t_{k-1}^{(N)}\rangle\to 1/(\gamma_a
k)$ as was observed in Refs.~\cite{wang08b, brune08a}. Now in
possession of the full probability distribution, we can also derive
statistical quantities previously out of reach even in the
limit~$\Gamma\to\infty$, such as the standard
deviation~$\sigma_{\Gamma,k}^{(N)}$ for the emission time of the~$k$th
photon from a~$N$-photon bundle,
$\gamma_a\sigma_{\infty,k}^{(N)}=
{\sqrt{H_{N,2}-H_{N-k,2}}}$~\cite{multisup}
with~$H_{N,2}\equiv\sum_{k=1}^N{k^{-2}}$ (generalized harmonic
numbers).  In the opposite limit, the same harmonic structure is found
but this time governed by the filter alone with no direct influence
(to leading order) from the radiative decay, with, e.g.,
$\langle t_k^{(N)}\rangle\sim(H_{N}-H_{N-k})/\Gamma$ as~$\Gamma\to0$,
so:
\begin{equation}
  \label{eq:Thu9Sep110413CEST2021}
  \langle\tau\rangle_N={H_{N-1}\over\Gamma}\text{ at small~$\Gamma$}
  \text{ and }{H_{N-1}\over\gamma_a}\text{ at large~$\Gamma$}\,.
\end{equation}
To unify these two simple and symmetric limits, one has to use
Eq.~(\ref{eq:Mon16Aug170254CEST2021}). This time-dilation of a
filtered bundle is shown on the floor of
Fig.~\ref{fig:Thu27May142405CEST2021}(b). We now turn to a richer,
more accessible and more insightful statistical quantity that will
further allow us to generalize our discussion to other types of
multiphoton emission, namely, the waiting-time distribution (wtd),
i.e., the probability distribution of the time difference between
successive photons.  For the SE of a two-photon bundle, this is
obtained from Eq.~(\ref{eq:Mon7Jun131211CEST2021}) as
$w_2(\tau)=\int_0^{\infty}\phi^{(2)}_\Gamma(t,\tau+t)\,dt$,
which~gives
$w_2(\tau)=
{\Gamma\gamma_a(\Gamma+\gamma)\over(\Gamma-\gamma_a)^2(3\Gamma+\gamma)(\Gamma+3\gamma_a)}(\Gamma(3\Gamma+\gamma_a)e^{-\gamma_a\tau}+\gamma_a(\Gamma+3\gamma_a)e^{-\Gamma\tau}-8\Gamma\gamma_ae^{-{\Gamma+\gamma_a\over2}\tau})$. This
is a tri-exponential decay that balances the radiative decay with its
filtering.  This quantity provides another way to the average time
length of a two-photon bundle as a function of filtering
% %
% \begin{equation}
%   \label{eq:Sun19Sep114819CEST2021}
%   \langle\tau\rangle_2=\int_0^\infty\tau w_2(\tau)
% d\tau={1\over\Gamma}+{1\over\gamma_a}+{2\over\gamma_a+\Gamma}-{9\over4(3\Gamma+\gamma_a)}-{9\over4(\Gamma+3\gamma_a)}
% \end{equation}
%
$\langle\tau\rangle_2=\int_0^\infty\tau w_2(\tau)
d\tau={1\over\Gamma}+{1\over\gamma_a}+{2\over\gamma_a+\Gamma}-{9\over4(3\Gamma+\gamma_a)}-{9\over4(\Gamma+3\gamma_a)}$
(this is, alternatively, obtained from
Eq.~(\ref{eq:Mon16Aug170254CEST2021}) as
$\langle t_2^{(2)}\rangle-\langle t_1^{(2)}\rangle$), which recovers
the limits~(\ref{eq:Thu9Sep110413CEST2021}) with numerator~$H_2=1$.
% %
% \begin{equation}
%   \label{eq:Thu9Sep115141CEST2021}
%   \langle t\rangle_2={1\over\Gamma}+{1\over\gamma_a}+{2\over\gamma_a+\Gamma}-{9\over4(3\Gamma+\gamma_a)}-{9\over4(\Gamma+3\gamma_a)}
% \end{equation}

\begin{figure}
  \includegraphics[width=\linewidth]{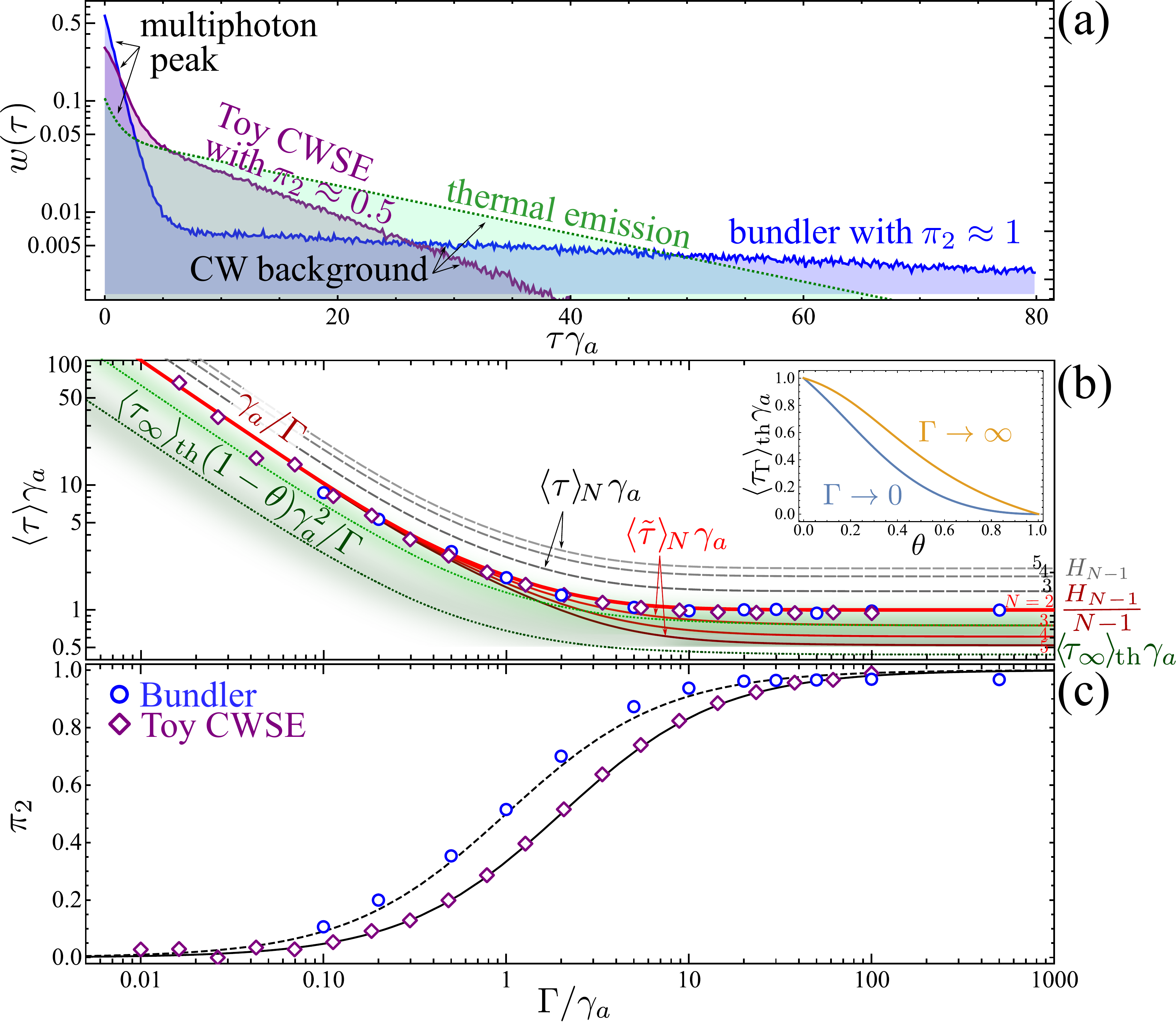}
  \caption{Dynamical multiphoton emission. (a) wtd for the toy-model
    (CWSE of Fock states, purple), the bundler (blue) and a thermal
    state (dotted green), all featuring a multiphoton peak sitting on
    the background of successive multiphotons. The toy model and
    bundler would be identical if they had the same purity.  (b)
    Multiphoton peak average as a function of filtering, for the case
    of pure multiphoton emission~$\langle\tilde\tau\rangle_N$ for~$N$
    from 2 to~5, exhibiting quantization at large~$\Gamma$ and a
    unique asymptote at small~$\Gamma$ when all bundles are broken
    into two-photon ones.  The toy-model (purple diamonds) and the
    bundler (blue circles) behave according to the SE picture (red
    thick line).  In contrast, the filtered thermal multiphoton-peak
    averages~$\langle\tau_\Gamma\rangle_\mathrm{th}$ are not quantized
    (shading) and retain multiphoton contributions to all orders.  In
    inset, average of the thermal multiphoton peak in its wtd as a
    function of $\theta\equiv P_a/\gamma_a$ leading to different
    asymptotes. Dashed gray lines: $N$-photon bundle time length
    $\langle\tau\rangle_N$ also featuring same coefficients for the
    asymptotes.  (c) Two-photon purity [percentage of fully detected
    bundles] from Fock state SE (line), the toy-model (diamonds) and
    the bundler (circles) which higher purity fits well with a SE rate
    of~$\approx\gamma_a/2$ (dotted lines). }
  \label{fig:Thu3Dec131932CET2020}
\end{figure}

This completes our description of filtered SE of Fock states, which is
the starting point and reference for all other types of multiphoton
emission, for instance, steady-state $N$-photon emission. We focus for
now on the simplest case~$N=2$. We simulate numerically with a
frequency-resolved Monte Carlo method~\cite{lopezcarreno18a} a
filtered ``bundler'' (emitter of bundles)~\cite{sanchezmunoz14a},
i.e., a coherently driven Jaynes-Cummings Hamiltonian in the
dispersive limit, that can also be understood as a Purcell-enhanced
Mollow-triplet.  We compare both its wtd and its photon-purity~$\pi_2$
(the percentage of unbroken bundles or photons emitted in pairs) with
a toy-model of CWSE (continuous wave spontaneous emission) that
simulates the actual emission of the bundler with randomly triggered
SE of two-photon Fock states.  For the most part, we find that the
multiphoton emission from the bundler behaves accordingly to the SE of
collapsed Fock states, as has been suggested all
along~\cite{strekalov14a}.  Their wtd, shown in
Fig.~\ref{fig:Thu3Dec131932CET2020}(a) for different purities (50\%
and 100\%), share the same qualitative main features. The multiphoton
peak, in particular, i.e., the short-time excess of probabilities due
to photons piling up from the multiphotons, behaves similarly under
filtering. This is shown in Fig.~\ref{fig:Thu3Dec131932CET2020}(b)
where numerically simulated CWSE and two-photon bundling are
superimposed on the theoretical~$\langle\tau\rangle_2$ above. There
are, nevertheless, noteworthy quantitative departures, most
prominently, the greater robustness of the bundles to filtering, as
seen in Fig.~\ref{fig:Thu3Dec131932CET2020}(c) where the purity
corresponds to that of SE with an effective decay rate
of~$\approx \gamma_a/2$. Therefore, for the same filter width, bundles
originating from the driven system (bundler) are significantly more
likely to be fully detected than if they were emitted by SE. This is
surprising given the otherwise excellent agreement with a picture of
collapsed Fock states that are subsequently spontaneously emitted. One
can track this departure to the dynamics of the emitter itself (the
two-level system) that, upon collapsing, resets the SE process in a
mechanism akin to the quantum Zeno effect~\cite{misra77a} that
effectively slows down the emission, i.e., reduces~$\gamma_a$. This,
in turn, makes the multiphoton detection more robust, since filtered
SE is indeed better for larger~$\Gamma/\gamma_a$.  This dynamical
enhancement depends on various parameters such as the statistics of
the bundles themselves, which warrants a study on its own. This
suggests, however, that the prospects of improving multiphoton
emission by frequency filtering are even better than has been
anticipated~\cite{sanchezmunoz18a}.

Finally, one cannot address multiphoton emission without considering
the most famous and pervading case, which has been foundational to
quantum optics~\cite{hanburybrown56b} and whose role remains central
for applications~\cite{valencia05a}, let alone this being the most
common type of light, namely, the thermal state.  This is obtained by
supplementing the Lindblad master equation of SE with a pumping term
$\frac{P_a}{2}(2 \ud{a} \rho {a} - a\ud{a} \rho - \rho a\ud{a})$
where~$P_a<\gamma_a$ drives the cavity to a steady
state~$\rho=(1-\theta)\sum_{n=0}^\infty\theta^n\ket{n}\bra{n}$ with
effective temperature~$\theta\equiv{P_a/\gamma_a}$. We can now precise
its perceived relationship to multiphoton emission by comparing
thermal light to pure multiphoton emission. The Laplace transform of
the wtd is obtained from the same transform~$\tilde g^{(2)}(s)$ of
the~$g^{(2)}(\tau)$ function~\cite{kim87a} as
$\tilde{w}_\mathrm{th}(s) =(1+1/[\gamma_a
n_a\tilde{g}^{(2)}(s)])^{-1}$ or, by inverse transform,
$w_{\mathrm{th}}(\tau)=\frac{2P_a
  \gamma_aQ_a^{-1}}{\gamma_a-P_a}\exp{\big[\frac{-(\gamma_a^2+P_a^2)\tau}{2(\gamma_a-P_a)}
  \big]} \big \lbrace Q_a \cosh \big[ \frac{Q_a
  \tau}{2(\gamma_a-P_a)}\big]\!-\!2P_a\gamma_a \sinh \big[ \frac{Q_a
  \tau}{2(\gamma_a-P_a)} \big] \big \rbrace$, where we have
defined~$Q_a^2\equiv P_a^4 -4P_a^3\gamma_a + 10 P_a^2\gamma_a^2 -4P_a
\gamma_a^3 + \gamma_a^4$. This has a similar shape
(cf.~Fig.~\ref{fig:Thu3Dec131932CET2020}(a), green dotted line) than
the wtd of the multiphoton emissions previously described, but behaves
distinctively under filtering. For instance, let us compare the
average of the thermal multiphoton
peak~$\langle\tau_\Gamma\rangle_\mathrm{th}$ with its pure multiphoton
counterpart in Eq.~(\ref{eq:Thu9Sep110413CEST2021}): in the limit of
large-bandwidth filtering, the average is that of a thermal state,
$\langle
\tau_\infty\rangle_\mathrm{th}=(\gamma_a-P_a)/(P_a^2+\gamma_a^2+Q_a)$,
shown in the inset of Fig.~\ref{fig:Thu3Dec131932CET2020}(b).  It is,
in units of~$\gamma_a$, a quantity between~0 (when~$P_a\to\gamma_a$
with the multiphoton peak becoming a Dirac~$\delta$ function) and~$1$
(when~$P_a\to0$ with the multiphoton peak reducing to two-photon
bunching and thus to the radiative lifetime). It is thus continuous,
in contrast to pure multiphoton emission that is quantized (through
the Harmonic numbers).  For finite~$\Gamma$, the average is not that
of a thermal state anymore, since, surprisingly, although filtering
has a tendency to thermalize quantum states, and does this exactly in
the limit~$\Gamma\to 0$ for all physical states, filtering a thermal
state does \emph{not} produce another thermal state, despite the fact
that all the Glaubler correlators of the filtered thermal
field~$g^{(n)}_\Gamma(0)=n!$ remain the same. Their time dynamics,
however, exhibit a different evolution~\cite{multisup}. Since in the
limit~$\Gamma\to 0$, a thermal state is again recovered as a universal
limit, one can obtain the asymptotic behaviour:
$\gamma_a\langle\tau_0\rangle_\mathrm{th}=\langle
\tau_\infty\rangle_\mathrm{th}(1-\theta)/\Gamma$. Filtering pure
multiphoton emission, on the other hand, produces bundles of various
sizes, which yields a multiphoton peak in the wtd
$\langle\tilde\tau\rangle_N\equiv{\sum_{k=2}^N{\langle
    t_k^{(N)}\rangle-\langle t_1^{(N)}\rangle\over
    k-1}P(k,N)/\sum_{k=2}^NP(k,N)}$ obtained by weighting the
contribution of each sub-bundle population. This is qualitatively
different from thermal emission, that retains thermal contributions
from multiphotons of all sizes.  Therefore, even if the wide-filter
asymptotes~$\langle\tau_\infty\rangle_\mathrm{th}$
and~$\langle\tilde\tau\rangle_N$ match, their narrow-filter trends
depart as shown in Fig.~\ref{fig:Thu3Dec131932CET2020}(b). This
disconnection of the two limits, as compared to
Eq.~(\ref{eq:Thu9Sep110413CEST2021}), shows that there are dynamical
features present in thermal equilibrium that break from the paradigm
of SE. In the limit~$\Gamma\to0$, another unexpected result is found:
the thermal field is monochromatic though chaotic, as was already
understood by Glauber~\cite{glauber63a}. The surprise is that its
temperature
$\theta_{\mathrm{th},\Gamma}=P_a\gamma_a/(P_a^2+(\gamma_a+\Gamma)(\gamma_a-P_a))$,
as compared to that of the unfiltered field
$\theta_{\mathrm{th}}=P_a/\gamma_a$, can be higher (if $\Gamma<P_a$),
although one is only interposing a passive element.  This apparent
paradox is due to both fields having different time scales, with the
filter slowing down times by delaying photons. As a result, the
filtered photons can have a flatter distribution of photon-numbers,
with higher probabilities to find more excited-states, but these also
have a longer lifetime, so are less likely to be emitted, thus
conserving energy while increasing temperature. This could also be
seen as a CW version of an equally surprising phenomenon when
subtracting a single photon from a thermal field, that results in
increasing its average population~\cite{parigi07a}. These various
findings in multiphoton emission provide insightful relationships and
departures between a thermal state and pure multiphoton sources.
Their closest encounter is at vanishing temperature where a thermal
state then behaves as a two-photon source but with most of its bundles
broken\dots\ A fairly subtle and elusive connection!

In conclusion, we have provided a comprehensive and analytical
description of the fundamental case of SE of Fock states, illustrated
with statistical observables of interest. The treatment through
frequency filtering allows to describe at a fundamental level the
detection process. While we have focused on time observables, the same
analysis could be done in the frequency domain by Fourier transform of
Eq.~(\ref{eq:Mon7Jun131211CEST2021}), providing probabilities of
detections at given frequencies.  Compared to other types of
multiphoton emission, we have shown that although a bundler behaves in
all respect as SE of collapsed Fock states, dynamical features exist
that protect bundles, making them significantly more resilient to
filtering than if they were generated by SE. This calls for further
studies to understand, characterize and exploit such dynamical
advantages. We have also highlighted connections and departures with
thermal light, which features multiphoton emission to all orders of a
different character than pure multiphoton emission with broken
bundles.  Such characterizations could also be made for superchaotic
light, superbunching, leapfrog processes in resonance fluorescence,
etc., which should enlighten on the nature, character and relationship
of these sources with pure multiphoton emission, thus allowing to
establish a zoology of its various types.

\begin{acknowledgements}
  FPL acknowledges support from Rosatom, responsible for the roadmap
  on quantum computing. EdV acknowledges the CAM Pricit Plan (Ayudas
  de Excelencia del Profesorado Universitario) and the TUM-IAS Hans
  Fischer Fellowship and projects AEI / 10.13039/501100011033
  (2DEnLight). CT acknowledges the Agencia Estatal de Investigaci\'on
  of Spain, under contract PID2020-113445GB-I00 and, with EdV, the
  Sinérgico CAM2020Y2020/TCS-6545 (NanoQuCo-CM).
\end{acknowledgements}

\bibliographystyle{naturemag}
\bibliography{sci,arXiv,multifi}

\end{document}

\flushbottom 

\title{Multiphoton Emission\\Supplementary Material}

\author{G.~D\'{i}az Camacho}
\thanks{These two authors contributed equally}
\affiliation{Departamento de F\'isica Te\'orica de la Materia
Condensada \& IFIMAC, Universidad Aut\'onoma de Madrid, 28049 Madrid,
Spain} 

\author{E.~Zubizarreta Casalengua}
\thanks{These two authors contributed equally}
\affiliation{Faculty of Science and
  Engineering, University of Wolverhampton, Wulfruna St, Wolverhampton
  WV1 1LY, UK}
\affiliation{Departamento de F\'isica Te\'orica de la Materia
Condensada \& IFIMAC, Universidad Aut\'onoma de Madrid, 28049 Madrid,
Spain} 

\author{J.~C.~{L\'{o}pez~Carre\~{n}o}}
\affiliation{Institute of Theoretical Physics, University of Warsaw, ul. Pasteura 5, 02-093, Warsaw, Poland
}
\author{S.~Khalid}
\affiliation{Faculty of Science and
  Engineering, University of Wolverhampton, Wulfruna St, Wolverhampton
  WV1 1LY, UK}

\author{C.~Tejedor}
\affiliation{Departamento de F\'isica Te\'orica de la Materia
Condensada \& IFIMAC, Universidad Aut\'onoma de Madrid, 28049 Madrid,
Spain} 

\author{E.~del~Valle}
\affiliation{Departamento de F\'isica Te\'orica de la Materia
Condensada \& IFIMAC, Universidad Aut\'onoma de Madrid, 28049 Madrid,
Spain} 
\affiliation{Institute for Advanced Study, Technical University of Munich, Lichtenbergstrasse 2a, D-85748 Garching, Germany.}
\affiliation{Faculty of Science and
  Engineering, University of Wolverhampton, Wulfruna St, Wolverhampton
  WV1 1LY, UK}

\author{F.~P.~Laussy}
\affiliation{Faculty of Science and 
  Engineering, University of Wolverhampton, Wulfruna St, Wolverhampton
  WV1 1LY, UK}
\affiliation{Russian Quantum Center, Novaya 100, 143025 Skolkovo,
  Moscow Region, Russia}
\email{fabrice.laussy@gmail.com}

\date{\today}

\begin{abstract}
  We provide details on the theoretical derivations which, although
  not necessary to the skilled Mathematician to derive the results in
  the text, may help in reproducing or extending them. In addition to
  exact but awkward combinatoric sums, we also provide explicit
  results---both for convenience and illustration---for all photons up
  to~$N=5$ or, for particularly voluminous expressions, $N=3$. We
  finally provide the spectrum and~$g^{(2)}(\tau)$ function of a
  filtered thermal field, so as to show that this is not a thermal
  field, featuring departures in its dynamics although it has the same
  density matrix.
\end{abstract}

\maketitle

\section{Joint probability distribution of SE multiphoton emission}

We derive~$\phi^{(N)}(\theta_1,\dots,\theta_N)$ the joint probability
distribution function (pdf) that the $k$th photon of a Fock
state~$\ket{N}$ at~$t=0$ in a freely-radiating cavity, is emitted at
time~$\theta_k$. To fix ideas with a concrete case, we also discuss in
parallel the case~$N=2$ for which we
compute~$\phi^{(2)}(\theta_1,\theta_2)$ that the first photon is
emitted at time~$\theta_1$ and the second photon is emitted at
time~$\theta_2$.  By construction, since the $k$th photon is emitted
after the $(k-1)$th and before the $(k+1)$th:
\begin{equation}
  \label{eq:Sat5Jun124642CEST2021}
  0\le \theta_1\le \theta_2\le\cdots\le \theta_N
\end{equation}
with~$\theta_N$ unbounded.  The distribution is defined on the
corresponding domain or, alternatively, the probability being zero
that a different order be observed, one can define the distribution
on~$\mathbb{R}^n$ but enforce the
time-ordering~(\ref{eq:Sat5Jun124642CEST2021}) by multiplying the
distribution with the support function
\begin{equation}
  \label{eq:Sat5Jun125012CEST2021}
  S(t_1,\dots,t_N)\equiv{\mathbbm{1}}_{[0, \theta_{2}[}(\theta_{1}){\mathbbm{1}}_{[\theta_1, \theta_{3}[}(\theta_{2})\cdots  {\mathbbm{1}}_{[\theta_{N-2}, \theta_{N}[}(\theta_{N-1})
{\mathbbm{1}}_{[\theta_{N-1}, \infty[}(\theta_{N})
\end{equation}
where~$\mathbbm{1}_T(t)$ is the indicator function which is~$1$
if~$t\in T$ and is~$0$ otherwise. For~$N=2$, the support
is~$S(t_1,t_2)=\mathbbm{1}_{[t_1,\infty[}(t_2)$ and restricts the
distribution to the upper-triangular part of the~$(t_1, t_2)$ space,
enforcing that the probability that the 1st photon comes after the 2nd
is zero. The integral of the probability distribution gives the
probability that a given number of photons have been detected up to
some time, which one can relate to the photon-counting formula~$P$
derived from Mandel's formula. Namely
\begin{equation}
  \label{eq:Sat5Jun125947CEST2021}
  P(t_1,\dots,t_N)=\int_0^{t_1}\cdots\int_0^{t_N}\phi^{(N)}(\theta_1,\dots,\theta_N)\,d\theta_1\cdots d\theta_N
\end{equation}
is the probability to have detected~$N$ photons, at the respective
times~$t_i$. We can invert Eq.~(\ref{eq:Sat5Jun125947CEST2021}) to
obtain the pdf as the derivatives of its primitive from the fundamental
theorem of calculus:
\begin{equation}
  \label{eq:Sun26Sep131256CEST2021}
  \phi^{(N)}(\theta_1,\dots,\theta_N)={\partial^N\over\partial t_1\cdots\partial t_N}P(t_1,\dots,t_N)\,.
\end{equation}
The photon-counting probability~$P(\{t_i\})\equiv P(t_1,\dots, t_N)$
is found in the main text from the Mandel photon-counting version
$P(n,T;N)$ of detecting~$n$ of them up to~$T$ for both filtered and
unfiltered Spontaneous Emission (SE). It is given by a binomial
distribution whose success parameter is the quantum
efficiency~$\mathscr{T}(T)$, or probability to detect a single photon
in SE between times~$0$ and~$T$. From~$P(n,T;N)$, and since the
structure of SE is that of Bernouilli trials in a sequence of
independent success (emission)/failure (no-emission) experiments, one
can get access to the probability of any multiphoton detection
configuration, e.g.,
$\mathscr{S}(t_1, t_2)\equiv\mathscr{T}(t_2)-\mathscr{T}(t_1)$ is the
probability to detect a single photon between times~$t_1$ and~$t_2$,
while the probability to detect one photon of a two-photon bundle,
between time~$0$ and~$t$ with the other not having yet been detected
is $2\mathscr{T}(t)(1-\mathscr{T}(t))$, and $\mathscr{T}(t)^2$ is the
probability to detect the two photons of a two-photon bundle anywhere
between times~$0$ and~$t$. In this way one can complete the link
between $P$ and~$\phi$. We also give an alternative, geometric
derivation for the case~$N=2$, with
$P(t_1,t_2)=\int_0^{t_1}\int_0^{t_2}\phi(\theta_1,\theta_2)\,d\theta_1
d\theta_2$
%
% \begin{equation}
%   \label{eq:Wed12May114305CEST2021}
%P(t_1,t_2)=\int_0^{t_1}\int_0^{t_2}\phi(\theta_1,\theta_2)\,d\theta_1 d\theta_2  
% \end{equation}
%
%
\begin{figure}
  \includegraphics[width=.5\linewidth]{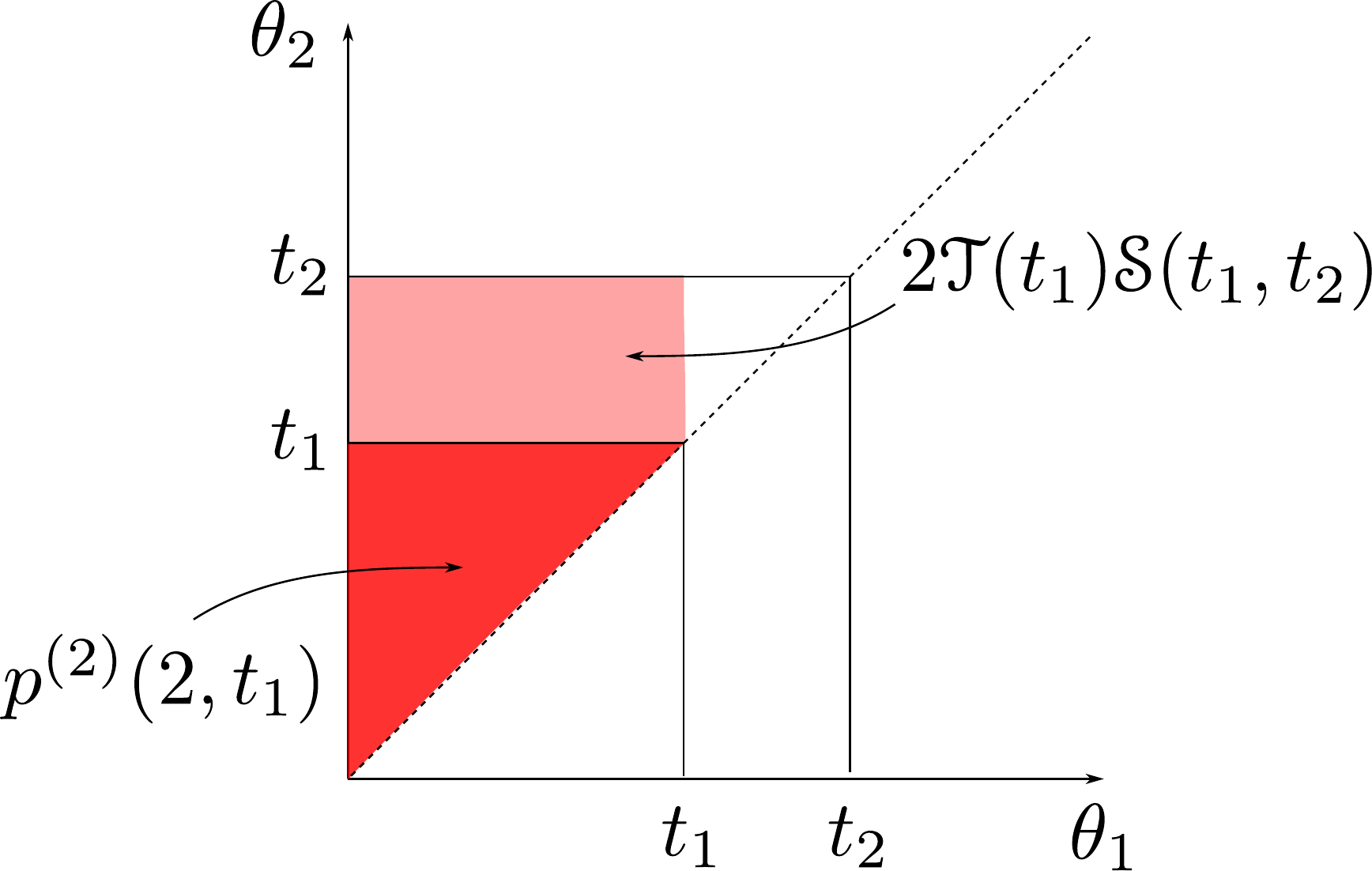}
  \caption{Two-photon probability space.}
  \label{fig:Sat5Jun130709CEST2021}
\end{figure}
being the probability to have detected two photons, the first one
between 0 and~$t_1$ and the second between 0 and~$t_2$ or, more
precisely and by construction, between $t_1$ and~$t_2$. One can
visualize this event in the 2D temporal plane as the reddish-areas
shown in Fig.~(\ref{fig:Sat5Jun130709CEST2021}), where the dark-red
triangle is the probability to detect the two photons in the time
window~$[0,t_1]$, i.e., $\mathscr{T}(t_1)^2$, and the upper,
lighter-red rectangle is the probability to detect the 1st photon up
to time~$t_1$ and the 2nd between time~$t_1$ and~$t_2$, i.e.,
${2\choose 1}\mathscr{T}(t_1)\mathscr{S}(t_1,t_2)$.  The lower
triangle is zero, by definition (1st photon comes first!) So the area
of interest integrates to a total probability of:
\begin{equation}
  \label{eq:Wed12May114126CEST2021}
  P(t_1,t_2)\equiv \mathscr{T}(t_1)^2+2\mathscr{T}(t_1)\mathscr{S}(t_1,t_2)\,.
\end{equation}
Applying Eq.~(\ref{eq:Sun26Sep131256CEST2021}) to this result,
i.e.,
$ \phi^{(2)}(t_1,t_2)={\partial^2\over\partial t_1\partial
  t_2}\int_0^{t_1}\int_0^{t_2}\phi^{(2)}(\theta_1,\theta_2)\,d\theta_1
d\theta_2$, derives the two-photon probability distribution from its
counting probabilities:
\begin{equation}
  \label{eq:Wed12May114333CEST2021}
  \phi^{(2)}(t_1,t_2)={\partial^2\over\partial t_1\partial t_2}\left(\mathscr{T}(t_1)^2+2\mathscr{T}(t_1)\mathscr{S}(t_1,t_2)\right)\,.
\end{equation}
Interestingly, the darker-red triangle, i.e., $\mathscr{T}(t_1)^2$,
does not contribute directly, which means that the two photons being
detected at exactly the same time is negligible as compared to other
possibilities.  The general case is not so easy to visualize 
geometrically but has otherwise the exact same formulation in an
hyperspace of photon time emissions which links~$\phi$ to~$P$ for~$N$
photons as:
\begin{equation}
  \label{eq:Mon7Jun095403CEST2021}
  \phi^{(N)}(t_1,\dots,t_n)={\partial^N\over\partial t_1\cdots\partial t_N}\left[{N\choose 1}\mathscr{T}(t)\prod_{k=1}^{N-1}\mathscr{S}(t_k,t_{k+1})\right]\,.
\end{equation}

Now it is just a matter to compute this expression for a given choice
of quantum efficiency~$\mathscr{T}$, that are given by Eqs.~(1)
and~(3) of the main text for unfiltered and for filtered SE,
respectively. We will now precise which case is which with a subscript.
In the first case (unfiltered SE or~$\Gamma\to\infty$), we find:
\begin{equation}
  \label{eq:Wed26May171858CEST2021}
  \phi^{(N)}_\infty(t_1,\dots,t_N)=N!\gamma_a^Ne^{-\gamma_a(t_1+\cdots+t_N)}\prod_{i=1}^{N}{\mathbbm{1}}_{[t_{i-1}, t_{i+1}[}(t_{i})
\end{equation}
while in the filtered case, we find:
\begin{equation}
  \label{eq:Mon7Jun131211CEST2021}
  \phi^{(N)}_\Gamma(t_1,\dots,t_N)=N! \gamma_a^N \left(\frac{\Gamma}{\Gamma_-}\right)^{2N }\prod_{i=1}^{N} (e^{-\Gamma t_i/2}-e^{-\gamma_a t_i/2})^2{\mathbbm{1}}_{[t_{i-1}, t_{i+1}[}(t_{i})
\end{equation}
where, in both cases, we have defined~$t_0\equiv 0$
and~$t_{N+1}\equiv\infty$. Equation~(\ref{eq:Mon7Jun131211CEST2021})
is Eq.~(4) of the main text. As should be,
Eq.~(\ref{eq:Wed26May171858CEST2021}) is a particular case
of~(\ref{eq:Mon7Jun131211CEST2021}), in the limit~$\Gamma\to\infty$
(whence the notation). Equation~(\ref{eq:Wed26May171858CEST2021})
makes particularly obvious the intrinsic symmetry of the emission
process: the photons are emitted independently the ones from the
others as far as the radiation mechanism is concerned, the joint pdf
being a simple product of their single-photon probabilities. It is the
domain restrictions
$\prod_{i=1}^{N}{\mathbbm{1}}_{[t_{i-1}, t_{i+1}[}(t_{i})$ that
correlate the photons in a nontrivial way, opening the radiation
window for the~$k$th photon when the~$(k-1)$th has been emitted and
closing it for all successives photons. The same structure is observed
for Eq.~(\ref{eq:Mon7Jun131211CEST2021}) which is a product of
one-dimensional distributions~$\varrho_\Gamma(t)$ with
\begin{equation}
  \label{eq:Mon7Jun131858CEST2021}
  \varrho_\Gamma(t)\equiv\sqrt[N]{N!}\gamma_a\left({\Gamma\over\Gamma_-}\right)^2(e^{-\Gamma t/2}-e^{-\gamma_at/2})^2
\end{equation}
(that reduces to
$\varrho_\infty(t)=\sqrt[N]{N!}\gamma_a e^{-\gamma_at}$ for unfiltered
emission). Equation~(\ref{eq:Mon7Jun131858CEST2021}) can thus be seen
as the (unnormalized) probability density of photon-detection for a
filtered SE photon.  There is a clear balance in this expression
between the limits where the filter and the radiation dominate the
dynamics, respectively, which leads to the two types of behaviour
captured, e.g., in Eq.~(6) of the main text. One could possibly, from
these results, postulate the general form of SE under any type of
detection once its single-photon response is known.

In addition to the impact of filtering on the temporal structure of
the emission, there is an even stronger effect, namely, that it
actually removes photons.  Because of this, the probability to detect
all the~$N$ photons from a filtered $N$-photon bundle emission is not
one. It is, instead, given by the integral of
Eq.~(\ref{eq:Mon7Jun131211CEST2021}) 
\begin{equation}
  \label{eq:Mon7Jun153112CEST2021}
  \mathcal{N}\equiv\idotsint_{0}^\infty\phi^{(N)}_\Gamma(t_1,\dots,t_N) dt_1\cdots dt_N=P(N,N)
\end{equation}
which is, as should be expected, the fraction of fully detected
$N$-photon bundles, providing the normalization
constant~$\mathcal{N}=(\Gamma/\Gamma_+)^N$ where we remind that
$\Gamma_{\pm} \equiv \Gamma \pm \gamma_a$. Depending on whether one is
interested in probabilities or in percentages, the distribution has to
be normalized, or not, as we will illustrate below on some particular
cases. The general case is given in the main text with~$\mathscr{T}$
given by Eq.~(3) in the limit of infinite times,
$P(k,N)\equiv\lim_{T\to\infty}p(k,T;N)={N\choose
  k}\big({\Gamma\over\Gamma_+}\big)^k\big(1-{\Gamma\over\Gamma_+}\big)^{N-k}$,
i.e.,
\begin{equation}
  \label{eq:Sun26Sep150830CEST2021}
%  P(k,N)\equiv{N\choose k}\left({\Gamma\over\Gamma_+}\right)^k\left(1-{\Gamma\over\Gamma_+}\right)^{N-k}\,.
  P(k,N)\equiv{N\choose k}{\gamma_a^{N-k}\Gamma^k\over\Gamma_+^N}\,.
\end{equation}
This gives the probability to detect~$k$ photons out of~$N$ from a
filtered $N$-photon bundle, and will be needed to normalize bundles
broken by filtering. We will turn, below, to the marginal
distributions~$\phi_{\Gamma,k}^{(N)}(t_k)$ for the~$k$th photon from
a~$N$-photon bundle to be emitted at time~$t_k$. Unfortunately, it
turns out that all the indices ($k, N, \Gamma$) are required to avoid
confusions in the notations. This is, by definition:
\begin{equation}
  \label{eq:Mon7Jun132048CEST2021}
  \phi_{\Gamma,k}^{(N)}(t_k)\equiv\idotsint_{0}^\infty\phi_\Gamma(\theta_1,\dots,\theta_N)\,d\theta_1\cdots d\theta_{k-1}d\theta_{k+1}\cdots d\theta_N\,,
\end{equation}
where the integral is taken over all but the~$k$th variable. %
% %\begin{equation}
% %  \label{eq:Sat12Jun103429CEST2021}
% \begin{multline}
%   \phi_\Gamma(t_1,\dots,t_N)=N! \gamma_a^N \left(\frac{\Gamma}{\Gamma_-}\right)^{2N }\\
%   (e^{-\Gamma t_1/2}-e^{-\gamma_a t_1/2})^2{\mathbbm{1}}_{[0, t_{2}[}(t_{1})(e^{-\Gamma t_2/2}-e^{-\gamma_a t_2/2})^2{\mathbbm{1}}_{[t_{1}, t_{3}[}(t_{2}) (e^{-\Gamma t_3/2}-e^{-\gamma_a t_3/2})^2{\mathbbm{1}}_{[t_{2}, t_{4}[}(t_{3})\cdots (e^{-\Gamma t_N/2}-e^{-\gamma_a t_N/2})^2{\mathbbm{1}}_{[t_{N-1}, \infty[}(t_{N})
% \end{multline}
% % \end{equation}
%
The calculation for the general filtered case, starting with
Eq.~(\ref{eq:Mon7Jun131211CEST2021}), proceeds as follows: integrating
first over~$t_N$ (the last time in the expression) reduces to the~1D
integral:
\begin{equation}
  \label{eq:Sat12Jun104950CEST2021}
  \int_{t_{N-1}}^\infty(e^{-\Gamma t_N/2}-e^{-\gamma_a t_N/2})^2dt_N={e^{-\gamma_a t_{N-1}}\over\gamma_a}+{e^{-\Gamma t_{N-1}}\over\Gamma}-4{e^{-\Gamma_+ t_{N-1}/2}\over\Gamma_+}
\end{equation}
where the previous photon only enters as the lower bound of
integration. Now integrating over the new last time in the expression,
$t_{N-1}$, we are met with a new~1D integral whose previous photon
(third to last) also enters as the lower bound, while the current
photon, although not the last, is left free to wander up to infinity
(there is enough room ``beyond infinity'' for the last photon to be
emitted after that):
\begin{equation}
  \label{eq:Sat12Jun145212CEST2021}
  \int_{t_{N-2}}^\infty(e^{-\Gamma t_{N-1}/2}-e^{-\gamma_a t_{N-1}/2})^2
\left({e^{-\gamma_a t_{N-1}}\over\gamma_a}+{e^{-\Gamma t_{N-1}}\over\Gamma}-4{e^{-\Gamma_+ t_{N-1}/2}\over\Gamma_+}\right)
%  e^{-\Gamma_+t_{N-1}}\left({e^{\gamma_a t_{N-1}}\over\Gamma}+{e^{\Gamma t_{N-1}}\over\gamma_a}-4{e^{\Gamma_+ t_{N-1}/2}\over\Gamma_+}\right)
dt_{N-1}\,.
\end{equation}
As merely a sum of exponentials, this is definitely integrable for any
particular case. To find the general case, we observe that
Eq.~(\ref{eq:Mon7Jun131211CEST2021}) can be rewritten as:
\begin{equation}
  \label{eq:Mon14Jun174617CEST2021}
  \phi_\Gamma(t_1,\dots,t_N)=(-1)^{N}N!\gamma_a^N\left(\frac{\Gamma}{\Gamma_-}\right)^{2N }\prod_{i=1}^{N} g'(t_i){\mathbbm{1}}_{[t_{i-1}, t_{i+1}[}(t_{i})
\end{equation}
where we defined
\begin{equation}
  \label{eq:Mon14Jun191900CEST2021}
  g(t)\equiv {e^{-\gamma_a t}\over\gamma_a}+{e^{-\Gamma t}\over\Gamma}-4{e^{-\Gamma_+ t/2}\over\Gamma_+}
\end{equation}
from Eq.~(\ref{eq:Sat12Jun104950CEST2021}), so that indeed
\begin{equation}
  \label{eq:Mon14Jun174453CEST2021}
  g'(t)=-(e^{-\Gamma t/2}-e^{-\gamma_a t/2})^2\,.
\end{equation}
The marginal distribution for the~$k$th photon is thus obtained as
(where the overline means ``skip this'' or ``don't perform''):
\begin{multline}
  \label{eq:Thu17Jun100949CEST2021}
  \phi_{\Gamma,\kappa}^{(N)}(t_\kappa)={(-1)^NN!\gamma_a^N}{\left({\Gamma\over\Gamma_-}\right)^{2N}}\times{}\\\int_{t_0}^{t_k}\int_{t_1}^{t_k}\cdots \int_{t_{k-2}}^{t_{k}}\overline{\int_{t_{k-1}}^\infty} \int_{t_{k}}^\infty \cdots \int_{t_{N-2}}^\infty\int_{t_N-1}^\infty g'(t_1)g'(t_2)\cdots g'(t_{k-1})g'(t_{k})g'(t_{k+1})\cdots{}\\{}\cdots g'(t_{N-1})g'(t_N) \,dt_1\cdots dt_{k-1}\overline{d_{t_k}}d_{t_{k+1}}\cdots dt_N
\end{multline}
which reduces the multi-dimensional integral to two series of
recursive~1D integrals of the type of
Eq.~(\ref{eq:Sat12Jun145212CEST2021}), one post and the other prior to
the $k$th photon emission. They both provide a functional dependence
in addition to the $k$th photon emission itself, $g'(t_k)$.  We
consider first the post-emission series. Integrating backward (last
photon first), we meet with the recurring pattern:
% $\kappa\equiv N-k$ with~$\kappa=0$
% %
% \begin{equation}
%   \label{eq:Mon14Jun171913CEST2021}
%   \int_{t_{N-(\kappa+1)}}^\infty g'(t_{N-\kappa})g^\kappa(t_{N-\kappa})\,dt_{N-\kappa}=-{g^{\kappa+1}(t_{N-(\kappa+1)})\over \kappa+1}
% \end{equation}
% %
%
\begin{equation}
  \label{eq:Mon14Jun171913CEST2021}
  \int_{t_{j-1}}^\infty g'(t_{j})g^{N-j}(t_{j})\,dt_{j}=-{g^{N-j+1}(t_{j-1})\over N-j+1}
\end{equation}
(since the $t\to\infty$ limit vanishes), whose recursive application
from the last~$j=N$ to the first post-emitted photon~$j=k+1$, yields:
\begin{equation}
  \label{eq:Wed16Jun140307CEST2021}
  \prod_{j=N}^{k+1}\int_{t_{j-1}}^{\infty}g'(t_j) dt_j={(-1)^{N-k}\over(N-k)!}g^{N-k}(t_k)
\end{equation}
where the decreasing product is to be performed in this order and
understood as implying nested integrals as shown in
Eq.~(\ref{eq:Thu17Jun100949CEST2021}). This $t_k$ dependence, from the
post-emission, factors out along with~$g'(t_k)$ the dynamical emission
of the~$k$th photon itself, and we are left to integrate the
pre-emission, for which we use the recurrence relation:
\begin{equation}
  \label{eq:Wed16Jun160241CEST2021}
  \int_{t_{i-1}}^{t_k}g'(t_{i})(g(t_{k})-g(t_{i}))^{n}\,dt_{i}={1\over n+1}(g(t_{k})-g(t_{i-1}))^{n+1}
\end{equation}
so that:
\begin{equation}
  \label{eq:Wed16Jun150321CEST2021}
  \prod_{i=k-1}^{1}\int_{t_{i-1}}^{t_k}g'(t_{i})\,dt_i={1\over(k-1)!}(g(t_k)-g(t_0))^{k-1}
\end{equation}
where the decreasing product is also to be understood as nesting
integrals. Bringing back the various pieces---the normalization
constant, Eq.~(\ref{eq:Wed16Jun140307CEST2021}), $g'(t_k)$
and~Eq.~(\ref{eq:Wed16Jun150321CEST2021})---together, we find:
\begin{equation}
  \label{eq:Thu17Jun114235CEST2021}
  \phi_{\Gamma,k}^{(N)}(t_k)=(-1)^NN!\gamma_a^N{\left({\Gamma\over\Gamma_-}\right)^{2N}}{(-1)^{N-k}\over(N-k)!}g^{N-k}(t_k)g'(t_k){1\over(k-1)!}(g(t_k)-g(t_0))^{k-1}
\end{equation}
which simplifies to the filtered one-photon marginal as:
\begin{equation}
  \label{eq:Thu17Jun114723CEST2021}
  \phi_{\Gamma,k}^{(N)}(t_k)=-{\left({\Gamma\over\Gamma_-}\right)^{2N}}k{N\choose k}\gamma_a^Ng(t_k)^{N-k}(g(t_0)-g(t_k))^{k-1}g'(t_k)\,.
\end{equation}
One could either reproduce the same calculation with
Eq.~(\ref{eq:Wed26May171858CEST2021}) instead of
Eq.~(\ref{eq:Mon7Jun131211CEST2021}), or take the
limit~$\Gamma\to\infty$, to obtain the marginal of the unfiltered
SE. In both cases, one finds the unfiltered probability distributions
of detecting the~$k$th photon as:
\begin{equation}
  \label{eq:lun12abr2021180549CEST}
  \phi_{\infty,k}^{(N)}(t_k)=
  k\gamma_a{N\choose k}e^{-N\gamma_a t_k}(e^{\gamma_a t_k}-1)^{k-1}\,.
\end{equation}
While Eq.~(\ref{eq:lun12abr2021180549CEST}) is normalized, for the
same reason as before, the marginals for the filtered emission,
Eq.~(\ref{eq:Thu17Jun114723CEST2021}) are not normalized to one but to
\begin{equation}
  \label{eq:Mon7Jun153842CEST2021}
  \int_0^\infty\phi_{\Gamma,k}^{(N)}(t)\,dt=P(N,N)
\end{equation}
which is the same result as Eq.~(\ref{eq:Mon7Jun153112CEST2021}). The
reason is that these marginals provide the probability distributions
when considering \emph{all} the~$N$ photons from a given photon
bundle, i.e., $\phi_{\Gamma,k}^{(N)}(t)$ is the probability to detect
the~$k$th photon from a $N$-photon bundle which has been detected in
its entirety (or conditioned to its full-detection). To take into
account that filtering occasionally removes some of the photons (any
number, from none to all), we introduce still another probability
distribution function, namely $\varphi_{\Gamma,k}^{(N)}$, which is the
probability density to detect the $k$th photon from a~$N$-photon
bundle of which any number from~$k$ to~$N$ photons have been
detected. For all~$N$,
$\phi_{\Gamma,N}^{(N)}=\varphi_{\Gamma,N}^{(N)}$ since a $N$th photon
is detected if and only if all photons have been detected.  The
general relationship is otherwise obtained from the law of total
probability
\begin{equation}
  \label{eq:Fri4Jun184130CEST2021}
  P(A_n)=\sum_kP(A_n|B_k)P(B_k)
\end{equation}
where we define
\begin{itemize}
\item $A_n$ as ``the detected photon is in the $n$th position of those detected'',
\item $B_k$ as ``there is a total of~$k$ photons detected out of the initial~$N$''
\end{itemize}
with the detection assumed to be in the interval~$[t,t+\Delta t[$, in
which case, since by definitions
$P(A_n)=\varphi_{\Gamma,n}^{(N)}(t)\Delta t$,
$P(A_n|B_k)=(\phi_{\Gamma,n}^{(k)}(t)/P(k,k))\Delta t$ (with~$P(k,k)$
the normalization of $\phi_n^{(k)}$ from
Eq.~(\ref{eq:Mon7Jun153842CEST2021})) and $P(B_k)=P(k,N)$, then
Eq.~(\ref{eq:Fri4Jun184130CEST2021}) becomes
$ \varphi_{\Gamma,n}^{(N)}=\sum_{k=1}^N{\phi_{\Gamma,n}^{(k)}\over
  P(k,k)}P(k,N)$ for all~$t$ and simplifying~$\Delta
t$. Since~$\phi_{\Gamma,n}^{(k)}$ is zero if~$n>k$, the sum can also
be taken from $k=n$ to~$N$. In this case, since $\cup_{k=n}^NB_k$ does
not form a partition of the space of possible detection events anymore
(the union from~$k=1$ to~$N$ does), therefore
$\varphi_{\Gamma,n}^{(N)}$ is not normalized to one but~to:
\begin{equation}
  \label{eq:Fri4Jun191414CEST2021}
  \mathcal{N}(n,N)\equiv\sum_{k=n}^NP(k,N)={\Gamma^n\gamma_a^{N-n}\over\Gamma_+^N}{N\choose n}{}_2F_1(1,n-N,n+1,-\Gamma/\gamma_a)
\end{equation}
which gives the fraction of detections with at least~$n$ photons in
the SE of $N$ photons, generalizing
Eq.~(\ref{eq:Mon7Jun153112CEST2021}) that was the particular
case~$\mathcal{N}(N,N)$. The result is also clear on physical grounds
as it states that any bundle which has been detected with exactly~$n$
photons contributes to all fractions with at least~$n$ photons. 
One can compute $P(k,N)/P(k,k)$ from
Eq.~(\ref{eq:Sun26Sep150830CEST2021}), so that the $k$th photon
normalized probability marginal from potentially broken bundles is
finally obtained~as
\begin{equation}
  \label{eq:Fri4Jun192029CEST2021}
  \varphi_{\Gamma,n}^{(N)}(t_n)={1\over \mathcal{N}(n,N)}\sum_{k=n}^{N}{N\choose k}\left(\frac{\gamma_a}{\Gamma_+}\right)^{N-k}\phi_{\Gamma,n}^{(k)}(t_n)\,.
\end{equation}

\begin{figure}
  \includegraphics[width=.5\linewidth]{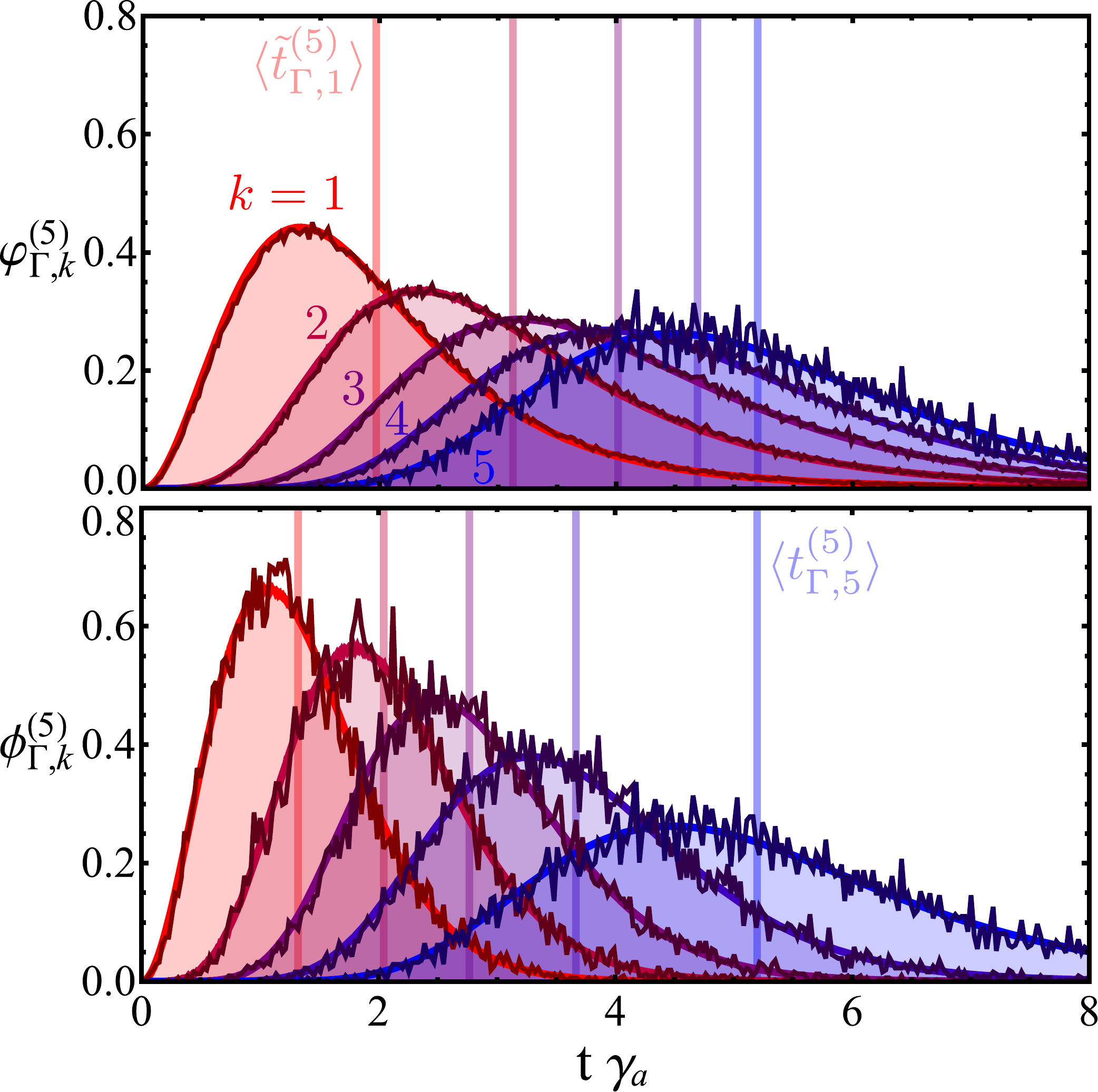}
  \caption{Reduced probability distributions (marginals) for the~$k$th
    photon of a~$(N=5)$-photon bundle conditioned to full-bundle
    dection~($\phi$, bottom panel) or measured on broken bundles due
    to filtering ($\varphi$, top panel), for the
    case~$\Gamma=\gamma_a$. The vertical lines show the average times,
    showing how filtering can reverse the piling-up effect of the
    bundles. Superimposed on the theoretical lines, Monte Carlo
    simulations confirm numerically the theoretical result.}
  \label{fig:Sun26Sep161621CEST2021}
\end{figure}

For instance, in the case shown in the main text for a five-photon
bundle, reproduced in more details in
Fig.~\ref{fig:Sun26Sep161621CEST2021}, one has for, say, the second
detected photon from all the possible detections:
\begin{equation}
  \varphi_{\Gamma,2}^{(5)}(t_2)\mathcal{N}(2,5)= \phi_{\Gamma,2}^{(5)}(t_2)+\frac{5\gamma_a}{\Gamma_+}\phi_{\Gamma,2}^{(4)}(t_2) +10\left( \frac{\gamma_a}{\Gamma_+}\right)^2\phi_{\Gamma,2}^{(3)}(t_2) +10\left( \frac{\gamma_a}{\Gamma_+}\right)^3\phi_{\Gamma,2}^{(2)}(t_2)
\end{equation}
with normalization
$\mathcal{N}(2,5)=\Gamma^2(\Gamma^3+5\Gamma^2\gamma_a+10\Gamma\gamma_a^2+10\gamma_a^3)/\Gamma_+^5$
the sum of the last four ($5-2+1$) terms in the binomial expansion
of~$\Gamma_+^5=(\gamma_a+\Gamma)^5$,
cf.~Eq.~(\ref{eq:Fri4Jun191414CEST2021}).  Such a result has also been
checked numerically with a Monte Carlo simulation that included
$N_\mathrm{traj}=400\,000$ photons, where ``traj'' stands for
``trajectory''. The corresponding data is also plotted in
Fig.~\ref{fig:Sun26Sep161621CEST2021}, which also illustrates how the
noise is distributed among the various cases. According to
Eq.~(\ref{eq:Fri4Jun191414CEST2021}), for~$\Gamma=\gamma_a$, the
expected numbers of fully-detected bundles is
$(\Gamma/\Gamma_+)^5N_\mathrm{traj}=N_\mathrm{traj}/32=12\,500$
(numerically $11\,542$ events have been recorded) while one expects
(resp.~finds) $62\,500$ bundles with one-missing photon (numerically
$59\,436$), $125\,000$ with two-missing (vs $122\,387$), $125\,000$
with three-missing (vs $126\,663$) and $62\,500$ with four-missing
photons, i.e., only one out of the five photons, has been detected
(numerically, $66\,093$). The difference from the number of attempts
goes with the completely missed bundles: $12\,500$ expected ($13\,909$
numerically). It is clear how, due to the smaller number of events,
the noise is greater for~$\phi_{\Gamma, k}$ as compared
to~$\varphi_{\Gamma, k}$, except for the case~$k=5$ where, not only
the amount, but the noise itself is the same in both cases
since~$\varphi_{\Gamma,5}^{(5)}=\phi_{\Gamma,5}^{(5)}$ and both
distributions are identically reconstructed from the available
signal. The amount of noise otherwise remains the same for lower~$k$
for~$\phi$ as all marginals are reconstructed from the same number of
events, $P(N,N)$, but decreases for~$\varphi$ since each event
with~$\kappa$ photons contributes to all $\varphi_{\Gamma,k}^{(N)}$
with~$k\ge\kappa$. Consequently, $\varphi_{\Gamma,1}^{(5)}$ has the
most signal for its reconstruction, and indeed displays an excellent
quantitative agreement with very little noise.

One can similarly derive multiphoton marginals. In the main text, we
focus in particular on the waiting time distribution (wtd) since it is
both popular and rich in information related to multiphoton emission,
particularly for two-photon emission which is the focus of the main
text. Without filtering, the SE of two-photons is trivial, with a wtd
$w_2(\tau)=\gamma_ae^{-\gamma_a\tau}$ that does not feature any
multiphoton effect, since the first-emitted photon, that gets a boost
in its decay time, sets the starting point for the next one, which is
just spontaneous emission. With filtering, however, the result becomes
sensibly richer:
\begin{subequations}
  \label{eq:Wed12May172337CEST2021}
  \begin{align}
    w_2(\tau)&={1\over P(2,2)}\int_0^{\infty}\phi^{(2)}_\Gamma(t_1,\tau+t_1)\,dt_1\\
             &={1\over P(2,2)}\int_0^\infty\varrho(t_1)\varrho(\tau+t_1)\,dt_1\\
             &={2\gamma_a^2\over P(2,2)}\left({\Gamma\over\Gamma_-}\right)^4\int_0^\infty[(e^{-\Gamma t_1/2}-e^{-\gamma_a t_1/2})(e^{-\Gamma(t_1+\tau)/2}-e^{-\gamma_a(t_1+\theta)/2})]^2\,dt_1\label{eq:Thu13May115526CEST2021}
  \end{align}
\end{subequations}
since $\mathbbm{1}_{[t_1,\infty[}(\theta+t_1)=1$. With the
normalization $P(2,2)=(\Gamma/\Gamma_+)^2$ to account for photon
losses from filtering, Eq.~(\ref{eq:Thu13May115526CEST2021}) gives:
\begin{equation}
  \label{eq:Thu13May115509CEST2021}
  w_2(\tau)={\Gamma\gamma_a(\Gamma+\gamma)\over(\Gamma-\gamma_a)^2(3\Gamma+\gamma)(\Gamma+3\gamma_a)}(\Gamma(3\Gamma+\gamma_a)e^{-\gamma_a\tau}+\gamma_a(\Gamma+3\gamma_a)e^{-\Gamma\tau}-8\Gamma\gamma_ae^{-{\Gamma+\gamma_a\over2}\tau})\,.
\end{equation}
This is the distribution of time differences of a two-photon bundle
that is spontaneously emitted and filtered through a filter of
width~$\Gamma$. That reduces to $\gamma_a e^{-\gamma_a\tau}$
for~$\Gamma\to\infty$ but for finite~$\Gamma$, this acquires the
tri-exponential decay form that bridges over the two limits of strong
and no filtering. Note that one-photon events, or broken bundles, do
not alter the result as the wtd is a two-photon observable. The
situation is different for higher~$N$ where photon losses should be
taken into account, or the observable conditioned to full-photon
detection. For instance, another two-photon marginal of interest is
the one that retains the first and last photon from a $N$-photon
bundle, tracing over the intermediate ones. This is useful to
characterize the statistics of observables such as the bundle
time-length. We find (some steps of the calculation will be detailed
below on other but similar cases):
\begin{multline}
  \label{eq:Sun24Oct123541CEST2021}
  \phi_\Gamma (t_1, t_N) = 2N!\left(\frac{\Gamma}{\Gamma_-}\right)^{2N}(-\gamma_a)^N \sum _{k_1+\cdots+k_6\atop = N-2} \sum_{k_7+\cdots+k_{10}\atop = 2}\prod_{i=1}^{10} \frac{1}{k_i!}\times{} \\
\frac{e^{-\frac{t_N}{2}(\gamma_a (2 k_1+k_7+k_8) +\Gamma (2 k_2 +k_9+k_{10}) +  \Gamma_+ k_3 )} e^{-\frac{t_1}{2}(\gamma_a (2 k_4 +k_7+k_9) +\Gamma (2 k_5 + k_8+k_{10}) +  \Gamma_+ k_6 )}}{ (-1)^{k_3+k_4+k_5+k_8+k_9} \gamma_a^{k_1+k_4} \Gamma^{k_2+k_5} \left(\frac{\Gamma_+}{4}\right)^{k_3+k_6}}\mathbbm{1}_{t_1\le t_N}\,.
\end{multline}

From the marginals, one can then compute statistical averages, for
instance the average times of detection, keeping in mind the
normalization Eq.~(\ref{eq:Mon7Jun153112CEST2021}). We work out some
cases in detail and otherwise list the results. We start with
$\langle t_1^{(N)}\rangle$, the average time of detection of the first
photon from a~$N$-photon bundle. This is obtained, by definition, as
$\langle t_1^{(N)}\rangle\equiv{1\over\mathcal{N}}\int_0^\infty
t_1\phi_{\Gamma,1}^{(N)}(t_1)\,dt_1$ and, from
Eq.~(\ref{eq:Thu17Jun114723CEST2021}), we have to compute
\begin{equation}
  \label{eq:Sun26Sep174328CEST2021}
  \langle t_1^{(N)}\rangle=-\left({\gamma_a\Gamma\Gamma_+\over\Gamma_-^2}\right)^NN\int_0^\infty t_1g(t_1)^{N-1}g'(t_1)dt_1
\end{equation}
which can be done by parts, since the integrand is $uv'$ where
  \begin{subequations}
    \begin{align}
      \label{eq:Sat19Jun103654CEST2021}
      u&=t & u'&=1 \\
      v&=(1/N)g(t)^N & v'&=g(t)^{N-1}g'(t)
    \end{align}
  \end{subequations}
  so that
  \begin{subequations}
    \label{eq:Sat19Jun103806CEST2021}
    \begin{align}
      \langle t_1^{(N)}\rangle&\propto uv\Big|_0^\infty-\int_0^\infty u'v\\
      &=-{1\over N}\int_0^\infty g(t)^N\,dt\,.\label{eq:Sun26Sep180520CEST2021}
    \end{align}
  \end{subequations}
  Given that $g(t)$ is a sum of three easily integrable terms,
  cf.~Eq.~(\ref{eq:Mon14Jun191900CEST2021}), we can use the
  multinomial theorem to find:
\begin{subequations}
  \label{eq:Sat19Jun101347CEST2021}
  \begin{align}
    \int_0^\infty g(t)^N\,dt&=\int_0^\infty\left({e^{-\gamma_a t}\over\gamma_a}+{e^{-\Gamma t}\over\Gamma}-4{e^{-\Gamma_+ t/2}\over\Gamma_+}\right)^N\,dt\\
                        &=\sum_{k_1+k_2+k_3=N}{N\choose k_1,k_2,k_3}\int_0^\infty{e^{-k_1\gamma_a t}\over \gamma_a^{k_1}}{e^{-k_2\Gamma t}\over\Gamma^{k_2}}{(-4)^{k_3}e^{-k_3\Gamma_+ t/2}\over\Gamma_+^{k_3}}\,dt\\
%                        &=\sum_{k_1+k_2+k_3=N}{(-4)^{k_3}{N\choose k_1,k_2,k_3}\over \gamma_a^{k_1}\Gamma^{k_2}\Gamma_+^{k_3}}\int_0^\infty{e^{-(k_1\gamma_a+k_2\Gamma+k_3\Gamma_+/2)t}}\,dt\\
%                        &=\sum_{k_1+k_2+k_3=N}{(-4)^{k_3}{N\choose k_1,k_2,k_3}\over \gamma_a^{k_1}\Gamma^{k_2}\Gamma_+^{k_3}}{1\over k_1\gamma_a+k_2\Gamma+k_3{\Gamma_+\over2}}\\
                        &=\sum_{k_1+k_2+k_3=N}{(-4)^{k_3}{N\choose k_1,k_2,k_3}\over \gamma_a^{k_1}\Gamma^{k_2}\Gamma_+^{k_3}(k_1\gamma_a+k_2\Gamma+k_3{\Gamma_+\over2})}\,,\label{eq:Thu14Oct154756CEST2021}
  \end{align}
\end{subequations}
The sum $k_1+k_2+k_3=N$ is taken over all of the $(N+1)(N+2)/2$
possible combinations of positive (including zero) integers that
satisfy this condition, e.g., for~$N=2$,
$(k_1,k_2,k_3)\in\{(0,0,2),(0,2,0),(2,0,0),(0,1,1),(1,1,0),(1,0,1)\}$. With
all pieces put together back again, the general expression for average
time of detection of the first photon from a $N$-photon bundle, reads:
\begin{equation}
  \label{eq:Fri22Oct094440CEST2021}
  \langle t_1^{(N)}\rangle=\left({\gamma_a\Gamma\Gamma_+\over\Gamma_-^2}\right)^NN\sum_{k_1+k_2+k_3=N}{(-4)^{k_3}{N\choose k_1,k_2,k_3}\over \gamma_a^{k_1}\Gamma^{k_2}\Gamma_+^{k_3}(k_1\gamma_a+k_2\Gamma+k_3{\Gamma_+\over2})}\,.
\end{equation}
Explicit cases for the first photon are given in
Eqs.~(\ref{eq:Tue16Nov103133CET2021}),
(\ref{eq:Sun26Sep180616CEST2021})
and~(\ref{eq:Sun26Sep180736CEST2021}) below. We now tackle the case
$k=N$, i.e., the last photon:
\begin{subequations}
  \label{eq:Sat19Jun095335CEST2021}
  \begin{align}
    \langle t_N^{(N)}\rangle\propto\int_0^\infty t_N(g(t_0)-g(t_N))^{N-1}g'(t_N)dt_N\\
    =uv\Big|_0^\infty-\int_0^\infty u'v
  \end{align}
\end{subequations}
where
\begin{subequations}
\begin{align}
  \label{eq:Sun13Jun100604CEST2021}
  u&=t & u'&=1 \\
  v&=(-1/N)(g(t_0)-g(t))^N & v'&=(g(t_0)-g(t))^{N-1}g'(t)\,.
\end{align}
\end{subequations}
So we now basically have to compute:
\begin{equation}
  \label{eq:Sat19Jun100706CEST2021}
  {(-1)^N\over N}\int_0^\infty g(t)^N\,dt
\end{equation}
since the $g(t_0)$ offset of the curve cancels in the calculation of
its area. Following the same procedured as detailed above, we arrive
to:

\begin{multline}
\label{eq:Thu14Oct154623CEST2021}
\langle t_N ^{(N)} \rangle =  N\left({\gamma_a\Gamma\Gamma_+\over\Gamma_-^2}\right)^N \sum_{k_1 + \cdots + k_6 =\atop N-1}{N-1 \choose k_1,k_2,k_3,k_4,k_5,k_6} \sum_{k_{7} + k_{8} = 2} {2\choose k_{7},k_{8}} \\ \frac{(-1)^{k_3 + k_4 + k_5 + k_{8}}}{\gamma_a^{k_1+k_4} \Gamma^{k_2+k_5} \left( \frac{\Gamma_+}{4} \right)^{k_3 + k_6} \big( \gamma_a (k_4+{k_{8}\over2}) +\Gamma(k_5 +{k_{7}\over2}) + \frac{\Gamma_+}{2} k_6\big)^2 }\,.
\end{multline} 

The general case for the $k$th photon from a $N$-photon bundle is
obtained similarly, and features a cancellation of the multinomial
terms which results in a fairly compact notation despite three
summations over the combinatorics:
\begin{multline}
  \label{eq:Sun26Sep213129CEST2021}
  \langle t_{k}^{(N)} \rangle = 2 N!\left({\gamma_a\Gamma\Gamma_+\over\Gamma_-^2}\right)^N\sum_{k_1 + k_2 + k_3\atop = N-k}\sum_{k_4 + \cdots+ k_9\atop = k-1} \sum_{k_{10} + k_{11}\atop = 2}\prod_{j=1}^{11}{1\over k_j!}\times\\ \frac{(-1)^{k_3 + k_6 + k_7 + k_8 + k_{11}}}{\gamma_a^{k_1 +k_4+k_7} \Gamma^{k_2+k_5+k_8} \left( \frac{\Gamma_+}{4} \right)^{k_3 + k_6 + k_9} \big( \gamma_a (k_1 + k_7+{k_{11}\over2}) +\Gamma( k_2 +k_8 +{k_{10}\over2}) + \frac{\Gamma_+}{2} (k_3+k_9)\big)^2 }\,.
\end{multline}
Summations are, as before (and from the multinomial formula), taken
over all positive integers that satisfy the condition and with
$\cdots$ indicating successive terms, i.e.,
$k_4+\cdots+k_9=\sum_{i=4}^9k_i$.
Equations~(\ref{eq:Fri22Oct094440CEST2021}) and
(\ref{eq:Thu14Oct154623CEST2021}) are, of course, recovered as
particular cases of Eq.~(\ref{eq:Sun26Sep213129CEST2021}), which is
the general result, Eq.~(5), retained in the main text.  From
combinatorics (stars and bars), there are $N+k-1\choose k-1$ ways to
add $k$ positive or zero integers to sum up to $N$. We give
explicitely the six expressions for~$\langle t_k^{(N)}\rangle$ from
for all~$k$ up to~$N=3$ (these expressions are also plotted in
Fig.~\ref{fig:Mon1Nov162120CET2021}):
\addtocounter{equation}{1}
\edef\eqSun26Sep180612CEST2021{\theequation}

\begin{equation}
  \label{eq:Tue16Nov103133CET2021}
\langle t_1^{(1)} \rangle = 
\frac{\gamma_a ^2+4 \gamma_a  \Gamma +\Gamma ^2}{\Gamma \gamma_a  (\Gamma +\gamma_a)}\,,\tag{{\eqSun26Sep180612CEST2021}a}
\end{equation}

\begin{equation*}
  \label{eq:Sun26Sep180616CEST2021}
  \langle t_1^{(2)} \rangle = \frac{3 \gamma_a^4 +31 \gamma_a^3 \Gamma +64\gamma_a^2\Gamma^2 +31 \gamma_a \Gamma^3 + 3 \Gamma^4}{2 \Gamma \gamma_a (\Gamma +\gamma_a) (\Gamma + 3 \gamma_a)(3 \Gamma + \gamma_a)}\,,\tag{{\eqSun26Sep180612CEST2021}b}
\end{equation*}

\begin{equation*}
  \label{eq:Sun26Sep180645CEST2021}
  \langle t_2^{(2)} \rangle = 
\frac{3 \left(3 \gamma_a ^4+19 \gamma_a ^3 \Gamma +40 \gamma_a ^2 \Gamma ^2+19 \gamma_a  \Gamma ^3+3 \Gamma ^4\right)}{2 \Gamma \gamma_a  (\gamma_a +\Gamma )^3 (3 \gamma_a +\Gamma ) (\gamma_a +3 \Gamma )}\,,\tag{{\eqSun26Sep180612CEST2021}c}
\end{equation*}

\begin{equation*}
  \label{eq:Sun26Sep180736CEST2021}
  \langle t_1^{(3)} \rangle = \frac{ 10 \gamma_a^6 +177 \gamma_a^5 \Gamma +800 \gamma_a^4 \Gamma^2 + 1298 \gamma_a^3 \Gamma^3 + 800 \gamma_a^2 \Gamma^4 + 177 \gamma_a \Gamma^5 + 10 \Gamma^6 )}{3 \Gamma \gamma_a (\Gamma +\gamma_a) (\Gamma + 2 \gamma_a) (2 \Gamma +\gamma_a) (5 \Gamma +\gamma_a) (\Gamma + 5\gamma_a)}\,,\tag{{\eqSun26Sep180612CEST2021}d}
\end{equation*}

\begin{equation*}
  \label{eq:Sun26Sep180758CEST2021}
  \langle t_2^{(3)} \rangle = \frac{\splitfrac{\Big(150 \gamma_a^8  + 2345 \gamma_a^7 \Gamma +14493 \gamma_a^6 \Gamma^2+ 41371 \gamma_a^5 \Gamma^3 +{}}{{} + 58786 \gamma_a^4 \Gamma^4 + 41371 \gamma_a^3 \Gamma^5 + 14493 \gamma_a^2 \Gamma^6 +2345 \gamma_a \Gamma^7 + 150 \Gamma^8\Big)}}{6 \Gamma \gamma_a (\Gamma +\gamma_a) (2 \Gamma + \gamma_a) (\Gamma + 2 \gamma_a)(3 \Gamma + \gamma_a)(\Gamma + 3 \gamma_a) (5 \Gamma +\gamma_a) (\Gamma + 5\gamma_a)}\,,\tag{{\eqSun26Sep180612CEST2021}e}
\end{equation*}

\begin{equation*}
\label{eq:Sun26Sep212538CEST2021}
\langle t_3^{(3)} \rangle = 
\frac{\splitfrac{ \Big(330 \gamma_a ^8+4511 \gamma_a ^7 \Gamma +23979 \gamma_a ^6 \Gamma ^2+65053 \gamma_a ^5 \Gamma ^3+{}}{{}+91918 \gamma_a ^4 \Gamma ^4+65053 \gamma_a ^3 \Gamma ^5+23979 \gamma_a ^2 \Gamma ^6+4511 \gamma_a  \Gamma ^7+330 \Gamma ^8\Big)}}{6 \Gamma \gamma_a  (\gamma_a +\Gamma) (2 \gamma_a +\Gamma ) (3 \gamma_a +\Gamma ) (5 \gamma_a +\Gamma ) (\gamma_a +2 \Gamma ) (\gamma_a +3 \Gamma ) (\gamma_a +5 \Gamma )}\,.\tag{{\eqSun26Sep180612CEST2021}f}
\end{equation*}
This shows how the complexity of these simple and fundamental
observables grows quickly for multiphotons. We could not find a
further-simplified expression for
Eq.~(\ref{eq:Sun26Sep213129CEST2021}) but do not exclude that it
exists. There are, indeed, alternative ways to write the particular
results, e.g., Eq.~(\ref{eq:Tue16Nov103133CET2021}) can also be
written as
$\langle t_1^{(1)} \rangle =
{1\over\gamma_a}+{1\over\Gamma}+{2\over\gamma_a+\Gamma}$ with an hoc
description that the time of emission of a filtered single photon
results from adding to its radiative time, the filtering time
$1/\Gamma$ and the time from their combined (averaged) emission rate
$1/({\gamma_a+\Gamma\over2})$. Such a forced reading of the equation
however becomes obscure for multiphotons, e.g.,
$ \langle t_1^{(2)} \rangle ={1\over
  8}\big({4\over\gamma_a}+{4\over\Gamma}+{8\over\gamma_a+\Gamma}+{9\over3\gamma_a+\Gamma}+{9\over\gamma_a+3\Gamma}\big)$. Similar
decompositions exist for other averages but we could not find them in
closed-form. We will discuss below how the unfiltered results,
however, simplify considerably.

We now compute, also for illustration that this can be done, other
statistical observables of possible interest. Considering, for
instance, the variance, one needs to find the squared arrival time for
the~$k$th photon from a $N$-photon bundle, which is:
\begin{multline}
  \label{eq:Thu14Oct162906CEST2021}
  \langle \big(t_k^{(N)}\big)^2\rangle  = 
  4 N! \left(\frac{\gamma_a\Gamma\Gamma_+}{\Gamma_-^2}\right)^N \sum_{k_1+k_2+k_3\atop=N-k}\sum_{k_4+\cdots+k_9\atop=k-1}\sum_{k_{10}+k_{11}\atop=2}\prod_{i=1}^{11}\frac{1}{k_i!}\times{}\\ \frac{(-1)^{k_3 + k_6 + k_7 + k_{8}+k_{11}}}{\gamma_a^{k_1+k_4+k_7} \Gamma^{k_2+k_5+k_8} \left( \frac{\Gamma_+}{4} \right)^{k_3 + k_6 +k_9} \big( \gamma_a (k_1+k_7+{k_{11}\over2}) +\Gamma(k_2+k_8+{k_{10}\over2}) + \frac{\Gamma_+}{2} (k_3+k_9)\big)^3}
\end{multline}
from which one obtains the standard deviation for the time of emission
as:
\begin{equation}
  \label{eq:Mon27Sep090951CEST2021}
  \sigma_{k}^{(N)}\equiv\sqrt{\langle \big(t_k^{(N)}\big)^2\rangle-\langle t_k^{(N)}\rangle^2}\,.
\end{equation}

Particular cases for the averaged squared times of detections are:
\begin{subequations}
  \label{eq:Thu14Oct214247CEST2021}
  \begin{align}
  \langle(t_1^{(1)})^2\rangle&= \frac{2 \left(\gamma_a^4+5 \gamma_a  \Gamma ^3 + 12 \gamma_a ^2 \Gamma ^2+5 \gamma_a^3 \Gamma +\Gamma ^4\right)}{\gamma_a^2 \Gamma ^2 (\gamma_a +\Gamma )^2}\,, \label{eq:Fri22Oct154656CEST2021}\\[.2cm]
  \langle(t_1^{(2)})^2\rangle&=\frac{\splitfrac{9 \gamma_a^8+132 \gamma_a^7 \Gamma+{}}{{} +886 \gamma_a ^6 \Gamma ^2+2636 \gamma_a ^5 \Gamma ^3+3810\gamma_a ^4 \Gamma ^4+2636 \gamma_a ^3 \Gamma ^5+886 \gamma_a ^2 \Gamma ^6+132 \gamma_a  \Gamma ^7+9 \Gamma ^8}}{2 \gamma_a ^2 \Gamma ^2 (\gamma_a +\Gamma )^2 (3 \gamma_a +\Gamma )^2 (\gamma_a +3 \Gamma )^2}\,,\label{eq:Fri22Oct175357CEST2021} \\
  \langle(t_2^{(2)})^2\rangle&= \frac{\splitfrac{63\gamma_a ^8 +708 \gamma_a ^7 \Gamma +{}}{{}+ 3322 \gamma_a ^6 \Gamma ^2+8684 \gamma_a ^5 \Gamma ^3+12462 \gamma_a ^4 \Gamma ^4+8684 \gamma_a ^3 \Gamma ^5+3322 \gamma_a ^2 \Gamma ^6+708 \gamma_a  \Gamma ^7+63 \Gamma ^8}}{2 \gamma_a ^2 \Gamma ^2 (\gamma_a +\Gamma )^2 (3 \gamma_a +\Gamma )^2 (\gamma_a +3 \Gamma )^2}\,,
  \end{align}
\end{subequations}
etc. Also in this case, one can find a decomposition in terms of
inverse squared rates for~$\langle(t_k^{(N)})^2\rangle$, e.g.,
$\langle(t_1^{(1)})^2\rangle/2={1\over\gamma_a^2}+{1\over\Gamma^2}+{3\over\gamma_a\Gamma}+{4\over(\gamma_a+\Gamma)^2}$
(cf.~Eq.~(\ref{eq:Fri22Oct154656CEST2021})) or
$4\langle(t_1^{(2)})^2\rangle={2\over\gamma_a^2}+{2\over\Gamma^2}+{12\over\gamma_a\Gamma}+{8\over(\gamma_a+\Gamma)^2}+{18\over(3\gamma_a+\Gamma)^2}+{18\over(\gamma_a+3\Gamma)^2}+{9\over\Gamma(3\gamma_a+\Gamma)}-{3\over\Gamma(\gamma_a+3\Gamma)}$
(cf.~Eq.~(\ref{eq:Fri22Oct175357CEST2021})), which again suggests
possible simplifications of the general
expression~(\ref{eq:Thu14Oct162906CEST2021}). The squared times in
themselves are not of great intrinsic interest, but the standard
deviations are.  We similarly provide them explicitly for the
detection times of all photons up to three-photon bundle. The
expressions are so bulky, however, that a notational device needs
being developed. Standard deviations are of the type, for the $k$th
photon of a $N$-photon bundle:
\begin{equation}
  \label{eq:Sat23Oct155425CEST2021}
    \sigma_k^{(N)}={\displaystyle\sqrt{\sum_{i=0}^{\mu_k^{(N)}}\alpha_{i,k}^{(N)}\gamma_a^i\Gamma^{\mu_k^{(N)}-i}}\over\displaystyle\gamma_a\Gamma\sum_{i=0}^{(\mu_k^{(N)}/2)-1}\beta_{i,k}^{(N)}\gamma_a^i\Gamma^{({\mu_k^{(N)}}/{2})-(i+1)}}
\end{equation}
where, for each~$k$, some constants~$\alpha_{i,k}^{(N)}$ for the
numerator and~$\beta_{i,k}^{(N)}$ for the denominator are defined
(with $\mu_k^{(N)}+1$ terms for the numerator and, to keep the
dimensionality correct, $\mu_k^{(N)}/2$ terms for the
denominator). For instance:
\begin{subequations}
  \label{eq:Sat23Oct180907CEST2021}
  \begin{align}
  \sigma_1^{(1)}&=\frac{\sqrt{\gamma_a ^4+2 \gamma_a ^3 \Gamma +6 \gamma_a ^2 \Gamma ^2+2 \gamma_a  \Gamma ^3+\Gamma ^4}}{\gamma_a \Gamma (\gamma_a+\Gamma)}\,,\label{eq:Thu18Nov171428CET2021}\\
    \sigma_1^{(2)}&= \frac{\sqrt{9 \gamma_a ^8 +78 \gamma_a ^7 \Gamma + 427 \gamma_a ^6 \Gamma ^2+1118 \gamma_a ^5 \Gamma ^3+1584 \gamma_a ^4 \Gamma ^4+1118 \gamma_a ^3 \Gamma ^5+427 \gamma_a ^2 \Gamma ^6+78 \gamma_a  \Gamma ^7+9 \Gamma ^8}}{2\gamma_a\Gamma \left(3 \gamma_a ^3+13 \gamma_a ^2 \Gamma +13 \gamma_a  \Gamma ^2+3 \Gamma ^3\right)}\,.
  \end{align}
\end{subequations}
By construction, $\alpha_{i,k}^{(N)}=\alpha_{N-i,k}^{(N)}$ for
all~$0\le i\le \mu_k^{(N)}$ and
$\beta_{i,k}^{(N)}=\beta_{N-i,k}^{(N)}$ for
all~$0\le i\le (\mu_k^{(N)}/2)-1$. It is therefore more concise to
only display the minimal set of required coefficients, for instance as
follows:
\begin{equation}
  \label{eq:Sat23Oct182456CEST2021}
    \sigma_k^{(N)}\to{(\alpha_{1,k}^{(N)},\, \alpha_{2,k}^{(N)},\, \cdots,\, \alpha_{(\mu_{k}^{(N)}-1)/2,k}^{(N)})\over{(\beta_{1,k}^{(N)},\, \beta_{2,k}^{(N)},\, \cdots,\, \beta_{(\mu_{k}^{(N)}-3)/4,k}^{(N)})}}\,.
\end{equation}
If a term can be factored out, then we write it in front of the
list. Since this structure holds for other statistical quantities,
e.g.,
Eqs.~(\hyperref[eq:Tue16Nov103133CET2021]{\eqSun26Sep180612CEST2021})
or Eqs.~(\ref{eq:Thu14Oct214247CEST2021}), we could have adopted this
convention for these quantities as well, and would higher
photon-number ever prove to be needed, such tabulations (similar to
Clebsch--Gordan tables in the problem of angular momentum) would
certainly be generalized. For the case of the standard deviation, the
coefficients grow so fast that this more concise notation is, this
time, mandatory. For all photons up to 3-photon bundles, the standard
deviations are found to be:
\begin{subequations}
  \label{eq:Sat23Oct120500CEST2021}
  \begin{align}
    \sigma_1^{(1)}&\to{(1,\, 2,\, 6)\over{(1)}}\,,\label{eq:Sat23Oct183929CEST2021}\\
    \sigma_1^{(2)}&\to{(9,\, 78,\, 427,\, 1\,118,\, 1\,584)\over2{(3,\,\,13)}}\,,\label{eq:Sat23Oct183958CEST2021}\\
    \sigma_2^{(2)}&\to{(45,\, 390,\, 1\,235,\, 2\,662,\, 3\,864)\over2{(3,\,\,13)}}\,,\\
    \sigma_1^{(3)}&\to{(100,\, 1740,\, 16\,609,\, 89\,638,\, 291\,672,\, 585\,918,\,738\,854)\over3(10,\, 87,\, 227)}\,,\\
    \sigma_2^{(3)}&\to{\splitfrac{(11\,700,\,281\,580,\,2\,927\,353,\,18\,667\,626,\, 82\,053\,985,\, 256\,611\,346,\,576\,542\,235,\,}{936\,554\,328,\, 1\,101\,372\,206)} \over6(30,\, 361,\, 1\,581,\, 3\,212)}\,,\\
    \sigma_3^{(3)}&\to{\splitfrac{(44\,100, 1\,061\,340, 11\,033\,869, 65\,375\,898, 250\,957\,345, 683\,779\,714, 1\,399\,513\,767,}{2\,177\,984\,712, 2\,534\,107\,982)} \over6(30,\, 361,\, 1\,581,\, 3\,212)}\,,
  \end{align}
\end{subequations}
where Eqs.~(\ref{eq:Sat23Oct183929CEST2021})
and~(\ref{eq:Sat23Oct183958CEST2021}) encode respectively the
expressions~(\ref{eq:Sat23Oct180907CEST2021}) written in full (this
should help clarify the notation). In these lists, for each $k$
and~$N$, $\mu_{k}^{(N)}$ is given by twice the numbers of terms in the
numerator minus one, i.e., five terms for $(1,\, 2,\, 6)$ (three terms
in the numerator of Eq.~(\ref{eq:Sat23Oct183929CEST2021}) so
$2\times 3-1$ terms in the numerator of
Eq.~(\ref{eq:Thu18Nov171428CET2021})); nine for
$(9,\, 78,\, 427,\, 1\,118,\, 1\,584)$, etc.  The resulting standard
deviations are shown as bands surrounding the means in
Fig.~\ref{fig:Mon1Nov162120CET2021}, scaled by $1/10$ to avoid
cluttering the plot. One must also keep in mind the log-scale in
reading the magnitude of the standard deviation. Closer inspection
shows, as can be seen on the figure, that the relative deviation,
$\sigma_k^{(N)}/\langle t_k^{(N)}\rangle$, is minimum
at~$\Gamma=\gamma_a$.

\begin{figure}
  \includegraphics[width=\linewidth]{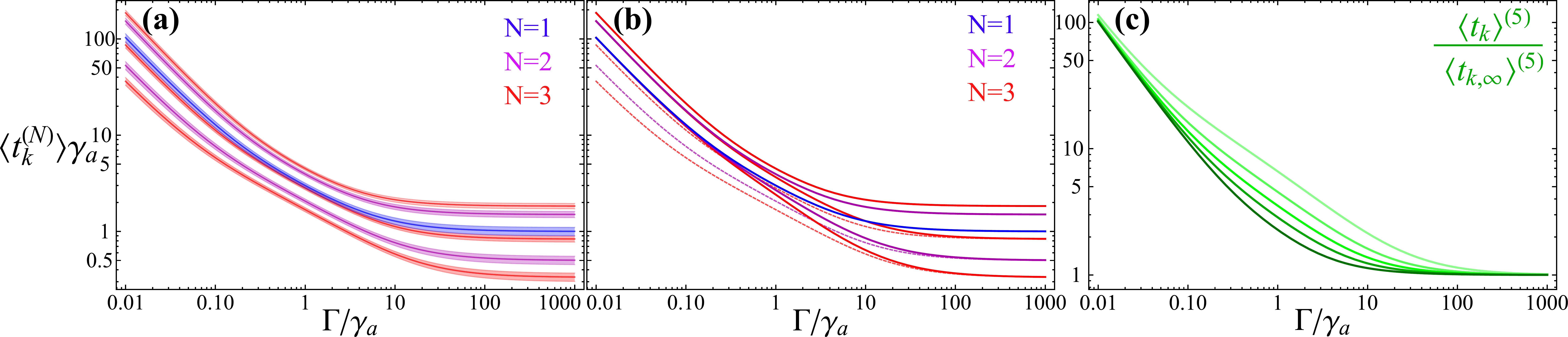}
  \caption{Mean time of detection~$\langle t_k^{(N)}\rangle$,
    Eqs.~(\hyperref[eq:Tue16Nov103133CET2021]{\eqSun26Sep180612CEST2021}),
    and its standard deviation ($\times1/10$),
    Eqs.~(\ref{eq:Sat23Oct120500CEST2021}), for every photon of a
    single-photon (blue), two-photon (pink) and three-photon bundle
    (red) as a function of filtering, conditioned to (a) full-photon
    detection, i.e., average is on~$\phi_{k,\Gamma}^{(N)}$ or to (b)
    broken-bundles due to filtering, i.e., average is on
    $\varphi_{k,\Gamma}^{(N)}$, without standard deviation for
    clarity. The unconditioned mean times from~(a) are also shown in
    dotted lines in~(b) for comparison. In~(c), the case~$N=5$ is
    shown with normalization to the asymptotes, to evidence the strong
    departures brought by filtering in the intermediate region.}
  \label{fig:Mon1Nov162120CET2021}
\end{figure}

One could carry on to higher photon-numbers, higher orders and/or for
other statistical quantities. It should be clear how to do this
following the examples that we have provided. Independently of whether
all such quantities will prove useful in the future, it is remarkable
how intricate even the simplest of them are for so fundamental
observables. Equation~(\ref{eq:Thu18Nov171428CET2021}), for instance,
is the standard deviation for the detection time of a single
photon. Single-photon emission is by far one of the most investigated
type of quantum light, with considerable in-depth analyses performed
over decades by countless groups worldwide. One would assume that the
standard deviation of its detection time would be well-known and
confirmed experimentally by now (from pulsed-emission, for
instance). It is, again, remarkable how such basic and fundamental
quantities have, however, resisted an exact description to date and
proved to be particularly complex.  We will turn to limiting cases
below, which are naturally much simpler.

By the symmetry discussed in the Main Text, visible in the expressions
(and spelled out to tabulate the coefficients for the standard
deviations~(\ref{eq:Sat23Oct120500CEST2021})), both $\Gamma\to\infty$
(no filtering) or $\Gamma\to 0$ (extreme filtering) display symmetric
behaviors, for which simple closed-form expressions can be given. The
general case might similarly admit such simplified versions, e.g.,
providing explicitly a formula for each~$\alpha_{i,k}^{(N)}$,
$\beta_{i,k}^{(N)}$, but we have not been able to find them in the
general case, although from the results above, one can provide any
particular case, for instance
$(\sigma_1^{(1)})^2={1\over\gamma_a^2}+{1\over\Gamma^2}+{4\over(\gamma_a+\Gamma)^2}$. In
particular, the corresponding coefficients, in one representation or
another, are not (yet) in the OEIS database of remarkable integer
sequences. Regardless of possible future simplifications and/or
enlightening interpretations, we wish to stress, however, that our
results already provide an exact and comprehensive description of
spontaneous emission of multiphotons. For instance, at the level of
averages and standard deviations on which we have focused above, the
second photon from a seven-photon bundle filtered with a
bandwidth~$\Gamma=\gamma_a$ is statistically detected at time
\begin{equation}
  \label{eq:Sat23Oct191630CEST2021}
  \gamma_a(\langle t_2^{(7)}\rangle\pm\sigma_2^{(7)})=\frac{554\,121\,805\,078\,044\,107}{325\,472\,664\,207\,527\,424}\pm\frac{\sqrt{36\,746\,843\,263\,868\,892\,857\,580\,028\,545\,723\,143}}{325\,472\,664\,207\,527\,424}
\end{equation}
which evaluates numerically to $\approx1.70251 \pm 0.588973$ (in units
of inverse~$\gamma_a$). The same exact treatment can be provided from
the above formulas for any particular instance and/or statistical
quantity.  Besides the theoretical appeal of closed-form descriptions,
it would not be surprising that one would wish to have such an exact
treatment for these fundamental quantities, which may ultimately be
involved in the operating details of, e.g., a photonic computer.

We finally provide, also both for illustration and for the intrinsic
interest of the statistical quantities derived, the computation of
averages from two-photon marginals. We have computed in
Eq.~(\ref{eq:Wed12May172337CEST2021}) the waiting time distribution
for a biphoton and in Eq.~(\ref{eq:Sun24Oct123541CEST2021}) the joint
first-and-last photons distributions. Corresponding two-photon
marginals include, e.g.,:
\begin{multline}
  \label{eq:Thu14Oct172435CEST2021}
  \langle t_1^{(N)} t_N^{(N)} \rangle =
  32 N!\left(-\frac{\gamma_a\Gamma\Gamma_+}{\Gamma_-^2}\right)^N \sum_{k_1+\cdots+k_6\atop = N-2} \sum_{k_7+\cdots+k_{10}\atop=2}\prod_{i=1}^{10}{1\over k_i!}{(-1)^{k_3+k_4+k_5+k_8+k_9}\over \gamma_a^{k_1+k_4} \Gamma^{k_2+k_5} \left(\frac{\Gamma_+}{4}\right)^{k_3+k_6}}\\
{}\times \frac{\gamma_a(6k_1+2k_4 + 4k_7+3k_8+k_9)+\Gamma(6k_2+2k_5+k_8+3 k_9+4 k_{10}) + \Gamma_+ (3k_3+k_6)} 
{ \splitfrac{\big(\gamma_a(2k_1+k_7+k_8)+\Gamma(2k_2+k_9+k_{10}) + \Gamma_+ k_3\big)^2 \times{}}{{}\times\big(\gamma_a(2k_1+2k_4+2k_7+k_8+k_9)+\Gamma(2k_2+2k_5+k_8+k_9+2k_{10}) + \Gamma_+ (k_3+k_6)\big)^3}}\,.
\end{multline}

\begin{figure}
  \includegraphics[width=.66\linewidth]{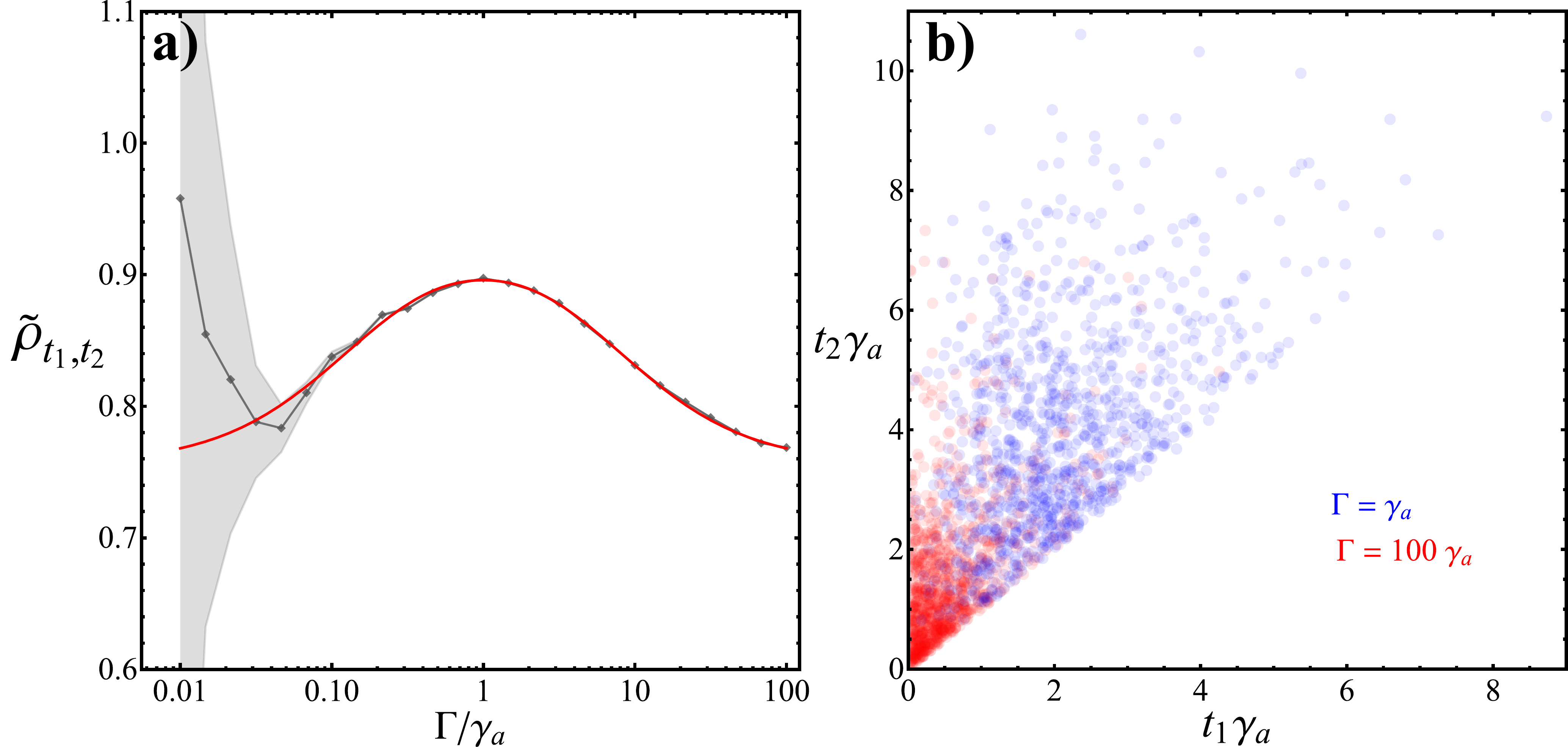}
  \caption{Two-photon emission correlation (a) as measured by the
    reflective Pearson
    correlator~Eq.~(\ref{eq:Sun24Oct114849CEST2021}) (theory, red, vs
    Monte Carlo, joined dots) and (b) as vizualized through a Monte
    Carlo scatter-plot. At small~$\Gamma$, the smaller available
    signal results in departures from the mean but these are due to
    exploding uncertainties (shaded area).}
  \label{fig:Mon18Oct102539CEST2021}
\end{figure}

This is a genuine multiphoton observable, that is valid for $N\ge 2$,
with, e.g.,
\begin{equation}
  \label{eq:Thu14Oct174041CEST2021}
    \langle t_1^{(2)} t_2^{(2)} \rangle = \frac{(\Gamma^2 + 4 \Gamma \gamma_a + \gamma_a^2)^2}{\Gamma^2 \Gamma_+^2\gamma_a^2}\,.
\end{equation}
The case~$N=1$ evaluates to zero due to the first sum running its
positive coefficients over a total of~$-1$, which is impossible. It
must thus be obtained directly from the single-photon marginal
Eq.~(\ref{eq:Thu14Oct162906CEST2021}).
For higher~$N$, this
allows us to compute, for instance, the covariance or related
statistical indicators, such as, to measure correlations, the
normalized covariance, known as the Pearson correlation coefficient:
\begin{equation}
  \label{eq:Thu14Oct175137CEST2021}
  \rho_{t_1,t_N}\equiv{\langle t_1^{(N)} t_N^{(N)} \rangle-\langle t_1^{(N)}\rangle\langle t_N^{(N)} \rangle\over\sqrt{\langle\big(t_1^{(N)}\big)^2\rangle-\langle t_1^{(N)}\rangle^2}\sqrt{\langle\big(t_N^{(N)}\big)^2\rangle-\langle t_N^{(N)}\rangle^2}}\,.
\end{equation}
For the case~$N=2$, the bulky expression shows an essentially constant
coefficient with exact limits~$\rho_{t_1,t_2}=1/\sqrt{5}\approx0.45$
at both $\Gamma\to0$ and~$\Gamma\to\infty$ with a slight increase to
$25/\sqrt{2929}\approx 0.46$ for~$\Gamma=\gamma_a$ where the
correlations are maximum. It can also be checked that these are
largely due to the geometrical constrain in the
distribution~(\ref{eq:Mon7Jun131211CEST2021}) imposed by the indicator
function, as previously commented. The Pearson correlator indeed
basically cancels if their ordering in time is randomly shuffled. On
the other hand, this is also impacted from the fact that the
distribution of events is not centered around their means. In this
case, a stronger indicator of correlation is given by the reflective
correlation coefficient, defined as:
\begin{equation}
  \label{eq:Sun24Oct114849CEST2021}
  \tilde\rho_{t_1,t_N}\equiv{\langle t_1^{(N)} t_N^{(N)} \rangle\over\sqrt{\langle\big(t_1^{(N)}\big)^2\rangle\langle\big(t_N^{(N)}\big)^2\rangle}}
\end{equation}
and, for~$N=2$, with both stronger correlations and stronger
variations of those correlations, from $2/\sqrt{7}\approx0.76$ at both
$\Gamma\to0$ and~$\Gamma\to\infty$ with a now sensibly greater
increase to $16/\sqrt{319}\approx 0.9$ for~$\Gamma=\gamma_a$ where the
correlations are also maximum. We do not discuss, but note, a
systematic increased degree of correlations between photons when
filtering them with a bandwidth that matches their radiative
rate. This is shown in Fig~\ref{fig:Mon18Oct102539CEST2021} for both
the theory and a Monte Carlo simulation. This precises one's physical
intuition of strong correlations but with no direct causality between
the events, i.e., if the first photon is emitted late in time, then
the second also has to, by construction, but without otherwise
actively influencing this from the radiating mechanism
itself. Correlations are smaller but remain sizable for the reflective
coefficient if shuffling the orders of arrival times. This means that
if one photon has been emitted late in time, chances are that the
other will also have been emitted late in time, again, with no
causality. The problem of photon correlations would be more
interesting for quantum correlated light, in which case these results
will provide a benchmark against which to evidence genuine
radiation-correlations.

\begin{figure}
  \includegraphics[width=.6\linewidth]{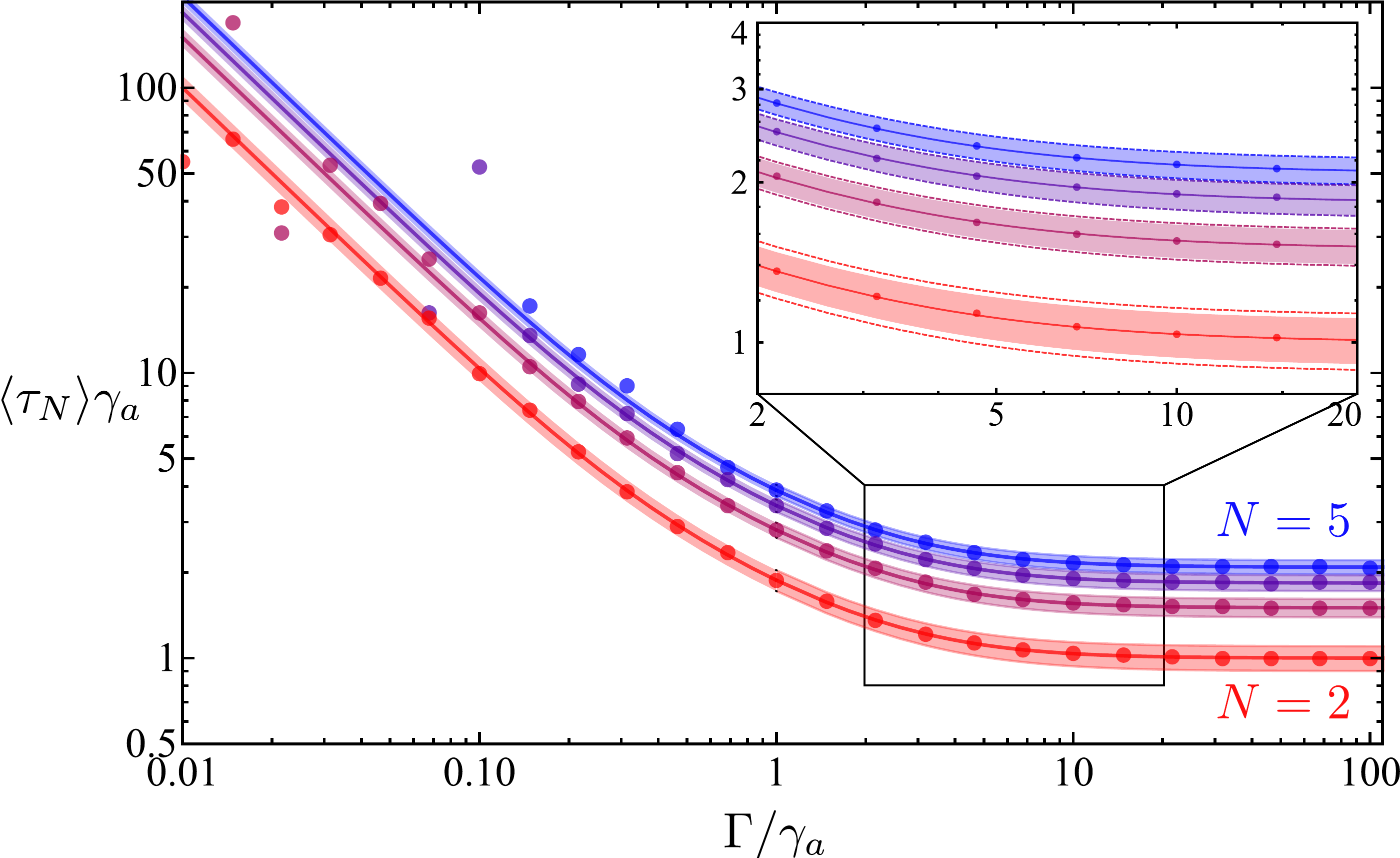}
  \caption{Average time length~$\langle\tau_N\rangle$ of a $N$-photon
    bundle ${}\pm\sigma_\tau^{(N)}/10$ for~$N=2$ (red) till~$5$
    (blue). Dots are Monte Carlo simulations, which depart from the
    exact theoretical result (lines) when~$\Gamma$ gets too small due
    to increasing numerical error caused by the small signal. In
    inset, a zoom shows the exact standard deviation (shaded) as
    compared to that obtained by
    assuming~$\langle t_1^{(N)}\rangle\langle t_N^{(N)}\rangle=\langle
    t_1^{(N)}t_N^{(N)}\rangle$, showing that the larger the~$N$, the
    smaller the photon correlations.}
  \label{fig:Sun31Oct193301CET2021}
\end{figure}

Also of interest is the statistics of the bundle time-length, i.e.,
over which period of time~$\tau_N$ does a $N$-photon bundle
extend. For the averages, there is no need of extra results, since,
defining:
\begin{equation}
  \label{eq:Sun24Oct125611CEST2021}
  \tau_N\equiv t_N^{(N)}-t_1^{(N)}
\end{equation}
the time difference between the last and first photons from the
bundles, then clearly
$\langle\tau_N\rangle=\langle t_N^{(N)}\rangle-\langle
t_1^{(N)}\rangle$ follows from Eqs.~(\ref{eq:Sun26Sep213129CEST2021}).
For a serious statistical treatment of the data, however, one would
also need the variance or standard deviation of
Eq.~(\ref{eq:Sun24Oct125611CEST2021}), which is
$\sigma_{\tau_N}^2\equiv\langle\tau_N^2\rangle-\langle\tau_N\rangle^2$,
i.e.,
\begin{equation}
  \label{eq:Sun24Oct130519CEST2021}
  \sigma_{\tau_N}^2=(\sigma_1^{(N)})^2+(\sigma_N^{(N)})^2+2\big(\langle t_1^{(N)}\rangle\langle t_N^{(N)}\rangle-\langle t_1^{(N)}t_N^{(N)}\rangle\big)
\end{equation}
which involves the covariance and thus requires
Eq.~(\ref{eq:Thu14Oct172435CEST2021}). We do not provide explicitly
the corresponding expressions, which can be obtained from the previous
results, but plot them up to~$N=5$ in
Fig.~\ref{fig:Sun31Oct193301CET2021}. From the various results that
can be discussed there, we will content to highlight that correlations
between the first and last photons (defining the bundle's duration or
time length) weaken with increasing~$N$, as could be expected. Also
interestingly, if normalized to their common asymptotes, unlike
Fig.~\ref{fig:Mon1Nov162120CET2021}(c) where there were strong
departures for the various~$k$, we find that the bundle lengths
depart very little from each others, namely $\tau_N/H_{N-1}$ differs
from~$\tau_2$ by less than 1\% of~$\tau_2$, for~$N$ up to~$5$ (one
needs~$N=6$ so that $(\tau_6/H_5-\tau_2)/\tau_2$ reaches an extremum
of~$\approx-1.12\%$). In all cases, the greatest departures always
occur at~$\Gamma=\gamma_a$.

We have seen how the general results involve complex combinatorics
interplay of the filter width and free radiation, which have been
expressed in exact closed-forms. Considerable simplifications are
obtained for the limiting cases. The unfiltered results, for instance
(which by symmetry are also those of vanishing filtering
widths~$\Gamma\to0$) are also of intrinsic and special interest. Their
general case has not been addressed either in the literature, to the
best of our knowledge, despite its tremendous fundamental
interest. They, too, can be obtained either directly from the
unfiltered pdf, Eq.~(\ref{eq:lun12abr2021180549CEST}), or by taking
the limit of the filtered result:
\begin{equation}
  \langle t_{\infty,k}^{(N)}\rangle\equiv \int_0^\infty t\phi_n^{(N)}(t)\,dt=\lim_{\Gamma\to\infty}\langle t_{k}^{(N)}\rangle\,.
\end{equation}
In both cases, we find:
\begin{equation}
  \label{eq:mar13abr2021002835CEST}
  \langle t_{\infty,k}^{(N)}\rangle={k\over\gamma_a}{N\choose k}
\sum_{l=0}^{k-1}(-1)^{k-1-l}{{k-1\choose l}\over(N-l)^2}\,.
\end{equation}
\begin{table}[]
\begin{tabular}{r|cccccccccc}
&1&&&&&\hskip-.5cm $k$&&&&10\\
\hline
1&1 &  &  &  &  &  &  &  &  &  \\ 
&${1\over2}$          & ${3\over2}$          &  &  &  &  &  &  &  &  \\[.1cm]
&${1\over3}$          & ${5\over6}$          & ${11\over6}$ &  &  &  &  &  &  &  \\[.1cm] 
&${1\over4}$          & ${7\over12}$          & ${13\over12}$ & $25\over12$ &  &  &  &  &  &  \\[.1cm] 
&${1\over5}$          & ${9\over20}$          & ${47\over60}$ & $77\over60$ & $137\over60$ &  &  &  &  &  \\%[.1cm] 
\raisebox{.2cm}{$N$}&${1\over6}$          & ${11\over30}$          & ${37\over60}$ & $19\over20$ & $29\over20$ & $49\over20$ &  &  &  &  \\[.1cm] 
&${1\over7}$          & ${13\over42}$          & ${107\over210}$ & $319\over420$ & $153\over140$ & $223\over140$ & $363\over140$ &  &  &  \\[.1cm] 
&${1\over8}$          & ${15\over56}$          & ${73\over168}$ & $533\over840$ & $743\over840$ & $341\over280$ & $481\over280$ & $761\over280$ &  &  \\[.1cm] 
&${1\over9}$          & ${17\over72}$          & ${191\over504}$ & $275\over504$ & $1879\over2520$ & $2509\over2520$ & $3349\over2520$ & $4609\over2520$ & $7129\over2520$ &  \\[.1cm] 
10&${1\over10}$         & ${19\over90}$          & ${121\over360}$ & $1207\over2520$ & $1627\over2520$ & $2131\over2520$ & $2761\over2520$ & $3601\over2520$ & $4861\over2520$ & $7381\over2520$ \\[.1cm] 
\end{tabular}
\caption{Values of $\langle t_{\infty,k}^{N}\rangle\gamma_a$,
  cf.~Eq.~(\ref{eq:mar13abr2021002835CEST}), up to~$N=10$. One can
  recognize several famous numerical series in these combinatorics,
  prominently, the reciprocals on the 1st column and the harmonic
  numbers on the diagonal. We show in the text how the intermediate
  values follow a simple pattern,
  cf. Eq.~(\ref{eq:Sun26Sep202753CEST2021})}
\label{tab:Sun26Sep192801CEST2021}
\end{table}
This result is a treasure trove of combinatoric, and the table of
coefficients that it gives rise to,
Table~\ref{tab:Sun26Sep192801CEST2021}, has both a rich structure and
a simpler one than is conveyed by
Eq.~(\ref{eq:mar13abr2021002835CEST}) which we can unravel from a
physical understanding of the nature of SE. In particular, the first
column is simply $\langle t_{\infty,1}^{(N)}\rangle=1/(N\gamma_a)$ as
is expected on physical grounds: the first photon to be emitted can be
any of the~$N$ available, each given an opportunity independent from
the others, thus speeding up its emission rate by a
factor~$N$. Similarly, the diagonal
$\langle t_{\infty,N}^{(N)}\rangle=H_N/\gamma_a$ is the time duration
or time-length of a bundle, and was already noted from the bundler's
emission as the result of subsequent SE in a cascade of emissions. The
second columns is formed, for its numerators, of the odd
numbers~$2N+1$, and for its denominators, of the oblong
numbers~$N(N+1)$, so that
$\langle t_{\infty,2}^{(N)}\rangle=(2N+1)/(N(N+1)\gamma_a)$. This is
better understood, however, as arising from the property
$\langle t_{\infty,2}^{(N)}\rangle-\langle
t_{\infty,1}^{(N)}\rangle=1/((N-1)\gamma_a)$ since once the first
photon has been emitted, there are~$N-1$ photons left and repeating
the same argument as before we arrive at the expected emission time
for this second photon. This argument can be iterated, so that one
expects:
\begin{equation}
  \label{eq:Sun26Sep200706CEST2021}
  \langle t_{\infty,k+1}^{(N)}\rangle-\langle t_{\infty,k}^{(N)}\rangle={1\over(N-k)\gamma_a}
\end{equation}
and, therefore, one can build Table~\ref{tab:Sun26Sep192801CEST2021}
by summing the Harmonic progression backward, summing from the
smallest number, e.g., for the~$N$th row: $1\over N$,
${1\over N}+{1\over N-1}$, ${1\over N}+{1\over N-1}+{1\over N-2}$,
\dots, $H_N$. This means that we can also
write~(\ref{eq:mar13abr2021002835CEST}) as $\langle t_{\infty,k}^{(N)}\rangle={1\over\gamma_a}\sum_{l=N-k+1}^N{1\over l}$ which simplifies to the physically
transparent form:
\begin{equation}
  \label{eq:Sun26Sep202753CEST2021}
  \langle t_{\infty,k}^{(N)}\rangle={H_N-H_{N-k}\over\gamma_a}\,.
\end{equation}
In particular:
\begin{subequations}
  \label{eq:Wed20Oct133235CEST2021}
  \begin{align}
  \langle t_{\infty,1}^{(N)}\rangle&={1\over N\gamma_a}\,,\\
  \langle t_{\infty,2}^{(N)}\rangle&={2N-1\over N(N-1)}{1\over\gamma_a}\,,\\
  &\cdots\nonumber\\
  \langle t_{\infty,N}^{(N)}\rangle&={H_N\over\gamma_a}\,.
  \end{align}
\end{subequations}

Once such a relationship between Eq.~(\ref{eq:mar13abr2021002835CEST})
and Eq.~(\ref{eq:Sun26Sep202753CEST2021}) is established, it is easy
to prove its mathematical validity for all cases by recurrence.
Equation~(\ref{eq:Sun26Sep202753CEST2021}) is remarkably simple
despite its rich structure. In this triangular tabulation, its
numerators are given by the OEIS sequence A213998 (numerators of the
triangle of fractions read by row) while its denominators are given by
the OEIS sequence A093919 (least-common multiple of integers listed in
reversed order). Again, while such simplifications or links to
combinatorics can be established in the limit~$\Gamma\to\infty$, we
could find no such description for the general case of
finite~$\Gamma$,
cf.~Eqs.~(\hyperref[eq:Tue16Nov103133CET2021]{\eqSun26Sep180612CEST2021}),
although it appears likely that similar reductions exist given the
occurences of remarkable sequences (one can spot, for instance,
Catalan numbers and number-theoretic partitions of integers, such as
the totient functions and greatest-common divisors). Those are mainly
considerations of elegance and aesthetic, however, which do not remove
to Eq.~(\ref{eq:Sun26Sep213129CEST2021}) its character of generality,
providing the exact result for all~$k$ and~$N$.

The same limits for the unfiltered emission of average squares
yields the general result 
\begin{equation}
  \label{eq:Mon27Sep092734CEST2021}
  \langle \big(t_{\infty,k}^{(N)}\big)^2\rangle=\frac{2 k}{\gamma_a^2} {N \choose k} \sum_{l=0}^{k-1}  (-1)^{k-1-l} \frac{{k-1 \choose l}}{(N-l)^3}
\end{equation}
from which the corresponding variances can be found. Some particular
cases can be obtained in a considerably simplified form, for instance:
\begin{subequations}
  \label{eq:Fri15Oct095455CEST2021}
  \begin{align}
    \langle \big(t_{\infty,1}^{(N)}\big)^2\rangle&={2\over(N\gamma_a)^2}\,,\\
    \langle \big(t_{\infty,2}^{(N)}\big)^2\rangle&=2{1+3N(N-1)\over\big(\gamma_a N(N-1)\big)^2}\,\\
    \cdots&\\
    \langle \big(t_{\infty,N}^{(N)}\big)^2\rangle&={1\over\gamma_a^2}\big(H_N^2+H_{N,2}\big)\,,
  \end{align}
\end{subequations}
where $H_{N,2}\equiv\sum_{k=1}^N{1\over k^2}$ is the 2nd generalized
harmonic number. As we shall now see, there is a contained closed-form
expression for this expression as well, namely:
\begin{equation}
  \label{eq:Wed20Oct141309CEST2021}
  \langle \big(t_{\infty,k}^{(N)}\big)^2\rangle={1\over\gamma_a^2}\left(H_{N,2}-H_{N-k,2}-(H_N-H_{N-k})^2\right)\,.
\end{equation}
This result can be inferred by inspection of the standard deviation
for the detection times of photons from unfiltered $N$-photon bundles,
obtained from Eqs.~(\ref{eq:Wed20Oct133235CEST2021})
and~(\ref{eq:Fri15Oct095455CEST2021}), for which one finds:
\begin{subequations}
  \label{eq:Mon27Sep093014CEST2021}
  \begin{align}
  \sigma_{\infty,1}^{(N)}&= \frac{1}{\gamma_a N} \\
  \sigma_{\infty,2}^{(N)}&= \frac{1}{\gamma_a}\sqrt{{1\over N^2}+{1\over(N-1)^2}} \\
    &\dots\nonumber\\
%  \sigma_{\infty,1}^{(3)}&= \frac{1}{3 \gamma_a} \\
%  \sigma_{\infty,2}^{(3)}&= \frac{\sqrt{13}}{6 \gamma_a} \\
  \sigma_{\infty,N}^{(N)}&= \frac{\sqrt{H_{N,2}}}{\gamma_a}\,.
\end{align}
\end{subequations}
This suggests, which can be confirmed a posteriori, that the contained
closed-form expression reads:
\begin{equation}
  \label{eq:Wed20Oct140947CEST2021}
  \sigma_{\infty,k}^{(N)}= \frac{\sqrt{H_{N,2}-H_{N-k,2}}}{\gamma_a}\,.
\end{equation}
From Eq.~(\ref{eq:Wed20Oct140947CEST2021}) and the definition of the
standard deviation~(\ref{eq:Mon27Sep090951CEST2021}), one can
therefore deduce Eq.~(\ref{eq:Fri15Oct095455CEST2021}) above. As
previously, once the result is found and checked for particular
values, it can be firmly established by recurrence.

Similarly, the statistics for the time-length of a fully-detected
photon bundle involves the computation of
$\langle (t^{(N)}_{\infty,N}-t^{(N)}_{\infty,1})^2\rangle$ which
itself requires:
\begin{equation}
  \label{eq:Sun24Oct190507CEST2021}
  \langle t_{\infty,1}^{(N)}t_{\infty,N}^{(N)}\rangle=
  \frac{(-1)^N}{\gamma_a^{2}} (N-1)\sum_{l=0}^{N-2} {N-2 \choose l} \frac{N+2l+2}{(-1)^{N-2-l}N^2(l+1)^2}
\end{equation}
which can be further simplified as:
\begin{equation}
  \label{eq:Mon1Nov155047CET2021}
  \langle t_{\infty,1}^{(N)}t_{\infty,N}^{(N)}\rangle={1\over\gamma_a^2}\left({H_N\over N}+{1\over N^2}\right)\,.
\end{equation}
From the above, one can now obtain the standard deviation for the
unfiltered bundle time length as the $\Gamma\to\infty$ limit of
Eq.~(\ref{eq:Sun24Oct130519CEST2021}):
\begin{equation}
  \label{eq:Mon27Sep093621CEST2021}
  \sigma_{\infty, \tau_N}^2=(\sigma_{\infty,1}^{(N)})^2+(\sigma_{\infty,N}^{(N)})^2+2\big(\langle t_{\infty,1}^{(N)}\rangle\langle t_{\infty,N}^{(N)}\rangle-\langle t_{\infty,1}^{(N)}t_{\infty,N}^{(N)}\rangle\big)
%  \sigma_{\infty,\tau}^{(N)}= \sqrt{\langle (t^{(N)}_{\infty,N}-t^{(N)}_{\infty,1})^2\rangle - \langle (t^{(N)}_{\infty,N}-t^{(N)}_{\infty,1})\rangle^2}
\end{equation}
which, from Eqs.~(\ref{eq:Sun26Sep202753CEST2021}),
(\ref{eq:Wed20Oct140947CEST2021}), and
(\ref{eq:Mon1Nov155047CET2021}), simplify to the simple general
result:
\begin{equation}
  \label{eq:Mon1Nov144635CET2021}
  % \sigma_\tau^{(n)}={1\over\gamma_a}\sqrt{\sum_{k=1}^n{1\over k^2}}
  \sigma_{\infty,\tau_N}={\sqrt{H_{N-1,2}}\over\gamma_a}
\end{equation}
i.e., $\sigma_{\infty,\tau_2}=\frac{1}{\gamma_a}$,
$\sigma_{\infty,\tau_3}= \frac{\sqrt{5}}{2 \gamma_a}$,
$\sigma_{\infty,\tau_4}= \frac{7}{6 \gamma_a}$,
$\sigma_{\infty,\tau_5}= \frac{\sqrt{205}}{12 \gamma_a}$, etc. That
summarizes nicely the statistics of $N$-photon bundles time lengths:
\begin{equation}
  \label{eq:Mon1Nov163336CET2021}
  \langle\tau_N\rangle\gamma_a=H_{N-1}\pm\sqrt{H_{{N-1},2}}\,.
\end{equation}

We believe that the above results provide a fairly comprehensive
overview of the statistics of bundle emission. What is not provided
explicitly can be either obtained from the general expressions or
obtained along similar lines of computation if not available here
(e.g., temporal correlations in broken bundles, skewness of emission
times, kurtosis, etc.) The unfiltered results happen to exhibit
beautiful connections to the generalized harmonic numbers. We can
imagine how still higher, $k$-orders statistical quantities would
likewise involve~$H_{N,k}$. We leave such generalizations for separate
investigations.

% \begin{equation}
%   \label{eq:Sat14Aug202730CEST2021}
%   \langle t_k^{(N)} \rangle =\left({\Gamma\over\Gamma_+}\right)^n{\sum_{i=0}^{\phi(N)}c_i\gamma^i\Gamma^{\phi(N)-i}\over \sum_\mathfrak{T}(a_i\gamma+b_i\Gamma)}
% \end{equation}
% %
% where $\mathfrak{S}$ is such that~$a_i+b_i\le 2N$ with $\gcd(a_i,b_i)=1$

% \begin{gather}
%   \label{eq:Mon16Aug170254CEST2021}\langle t_{k}^{(N)} \rangle = k \gamma_a^N \left( \frac{\Gamma}{\Gamma_-} \right) ^{2N} {N\choose k} \sum_{\kappa_{1,2,3}\atop= N-k} {N-k\choose k_1, k_2, k_3} \\\sum_{\kappa_{4,5,6,7,8,9}\atop = k-1}{k-1 \choose k_4,k_5,k_6,k_7,k_8,k_9} \sum_{k_{10} + k_{11} = 2} {2\choose k_{10},k_{11}}\frac{(-1)^{k_3 + k_6 + k_7 + k_8 + k_{11}}}{\gamma_a^{k_1 +k_4+k_7} \Gamma^{k_2+k_5+k_8} \left( \frac{\Gamma_+}{4} \right)^{k_3 + k_6 + k_9} ( \gamma_a (k_1 + k_7+k_{11}/2) +\Gamma( k_2 +k_8 +k_{10}/2) + \frac{\Gamma_+}{2} (k_3+k_9))^2 }
% \end{gather}
% %
% where we defined:
% %
% \begin{equation}
%   \label{eq:Mon16Aug170545CEST2021}
%   \kappa_{i,j,\cdots,l}\equiv k_1+k_j+\cdots+k_l
% \end{equation}

\section{``Thermalizing'' a thermal state}

We write ``thermalizing'' to describe the effect of filtering because
a filter that is narrow enough provides a thermal state out of any
quantum field, even single-photon
emission~\cite{delvalle12a,gonzaleztudela13a}. The only exception is
if the field has unphysical attributes, e.g., a $c$-number laser field
with no spectral width~\cite{gonzaleztudela13a}. It is therefore a bit
surprising that filtering a thermal field does not produce another
thermal field, except, of course, in the limit of a vanishing filter
bandwidth since this applies to all fields. This is even more
surprising given that all the Glauber correlators at zero delay remain
those of a thermal field, namely:
\begin{equation}
  \label{eq:Tue21Sep114423CEST2021}
  g^{(n)}(0)=n!\quad\text{for all~$n\ge 2$.}
\end{equation}
The field remains thus closely related to a thermal field in any case,
but with some important departures in the finite time correlations,
i.e., the $g^{(n)}(\tau)$, including~$g^{(1)}(\tau)$ whose Fourier
transform provides the normalized power spectrum:
\begin{equation}
  \label{eq:Sat25Sep102105CEST2021}
  S_{\mathrm{th},\Gamma}(\omega)={\pi\over2}{S_{\mathrm{th}}(\omega)}S_\Gamma(\omega)(\Gamma+\gamma_a-P_a)
\end{equation}
where~$S_{\mathrm{th}}(\omega)$ and~$S_{\Gamma}(\omega)$ are the
nomalized spectral shapes of the thermal state and of the filters,
respectively:
\begin{subequations}
  \label{eq:Sat25Sep120908CEST2021}
  \begin{align}
    S_{\mathrm{th}}(\omega)&={1\over\pi}{(\gamma_a-P_a)/2\over\big((\gamma_a-P_a)/2\big)^2+\omega^2}\,,\\
    S_{\Gamma}(\omega)&={1\over\pi}{\Gamma/2\over(\Gamma/2)^2+\omega^2}\,.
  \end{align}
\end{subequations}

\begin{figure}
  \includegraphics[width=\linewidth]{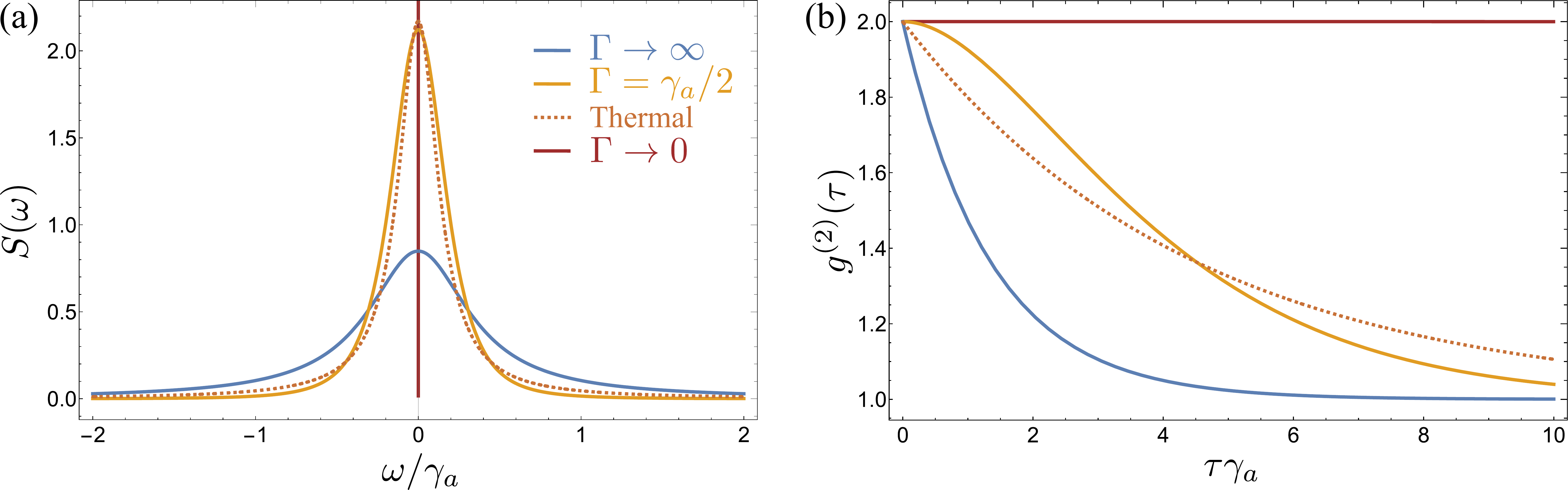}
  \caption{A thermal field of effective temperature~$\theta=0.25$
    with~$P_a=\theta\gamma_a$ (blue) as seen through its power
    spectrum (a) and second-order correlation function (b), as
    compared to its filtered field with a filter of
    width~$\Gamma=\gamma_a/2$ (yellow) and in the limit of vanishing
    filtering~$\Gamma\to 0$ (dark red). In the latter case, the
    spectrum becomes a Dirac~$\delta$ function while the correlation
    becomes constant. The dotted line shows the best-fit of the
    filtered thermal field by an actual thermal field, showing that
    the spectrum is not Lorentzian and that the decay of the
    correlation function is of a wholly different character. The
    limiting case~$\Gamma\to0$ can recover, however, an exact thermal
    field, by brushing off these departures to infinity.}
  \label{fig:Sat25Sep151950CEST2021}
\end{figure}

This shows how a thermal field, created under incoherent driving (at
rate~$P_a$) and spontaneous decay (at rate~$\gamma_a$), that is
filtered with a Lorentzian filter of bandwidth~$\Gamma$, does
\emph{not} itself have a Lorentzian spectral line, and thus cannot be
the result of a thermal equilibrium in a cavity.  Instead, filtering
produces a spectrum with less-fat tails than the thermal field itself,
as shown in Fig.~\ref{fig:Sat25Sep151950CEST2021}(a) where the
narrower field, thanks to filtering, is, however, non-Lorentzian. To
the best of our knowledge, this spectral shape has not reported in the
literature, despite its importance prior to the laser and the modern
theory of optical coherence, since this was the privileged way to
produce a monochromatic field: by filtering thermal~light.

We also give the second-order correlation of the filtered thermal
field to show how these departures take place also for higher-order
correlators:
\begin{equation}
  \label{eq:Sat25Sep104950CEST2021}
  g_{\mathrm{th},\Gamma}^{(2)}(\tau)=1+{1\over(\Gamma-\gamma_a+P_a)^2}\left(\Gamma^2e^{-(\gamma_a-P_a)\tau}+(\gamma_a-P_a)^2e^{-\Gamma\tau}-2\Gamma(\gamma_a-P_a)e^{-(\Gamma+\gamma_a-P_a){\tau/2}}\right)\,.
\end{equation}
This shows that although~$g_{\mathrm{th},\Gamma}^{(2)}(0)=2$, the
subsequent~$\tau$ dynamics differs from that of a thermal field, that
is the one recovered when filtering at all frequencies:
\begin{equation}
  \label{eq:Sat25Sep113257CEST2021}
  g^{(2)}_\mathrm{th}(\tau)=\lim_{\Gamma\to\infty}g_{\mathrm{th},\Gamma}^{(2)}(\tau)=1+\exp\big(-(\gamma_a-P_a)\tau\big)\,.
\end{equation}
The qualitative difference for this quantity with the two types of
fields is shown in Fig.~\ref{fig:Sat25Sep151950CEST2021}(b). The other
limit is also interesting:
\begin{equation}
  \label{eq:Sat25Sep115415CEST2021}
    g^{(2)}_\mathrm{th,0}(\tau)=\lim_{\Gamma\to0}g_{\mathrm{th},\Gamma}^{(2)}(\tau)=2\,.
\end{equation}
We postultate that~$g^{(n)}_\mathrm{th,0}(\tau)=n!$ for all~$n$. The
correlations are constant because in this limit, the filter vanishes
faster than the system's inverse coherence time, as seen in
Eq.~(\ref{eq:Sat25Sep113257CEST2021}). The power spectrum also
acquires an extreme form in this limit, the only one that allows it to
qualify as an exact thermal state:
\begin{equation}
  \label{eq:Sat25Sep121211CEST2021}
  S_{\mathrm{th},0}(\omega)\equiv\lim_{\Gamma\to0}S_{\mathrm{th},\Gamma}(\omega)=\delta(\omega)\,.
\end{equation}
The drawback is, of course, that the signal itself is zero. For
finite~$\Gamma$, one indeed obtains the emission rate of the filtered
thermal field as
$\gamma_a P_a\Gamma/[(\gamma_a-P_a+\Gamma)(\gamma_a-P_a)]$ that is
always smaller than the unfiltered ($\Gamma\to\infty$) rate~$I_\mathrm{th}$,
as should be by conservation of energy, and decays linearly with
filtering with a small enough~$\Gamma$,
$I_{\mathrm{th},\Gamma}\approx I_\mathrm{th}\Gamma$. Ignoring
the~$\tau$ dynamics, that becomes constant in this limit, since the
density matrix itself is that of a thermal state, and the dynamics is
``frozen'', one can speak of the effective temperature for the
filtered thermal field, with the interesting result that it can
actually be higher than that of the original field. Indeed, the
filtered-field temperature is
$\theta_{\mathrm{th},\Gamma}=P_a\gamma_a/(P_a^2+(\gamma_a+\Gamma)(\gamma_a-P_a))$
as compared to that of the unfiltered field
$\theta_{\mathrm{th}}=P_a/\gamma_a$. One can see how, when
$\Gamma<P_a$, the filtered field thus has a higher temperature than
the field that it is filtering. This apparent paradox is explained in
the main text.

\bibliographystyle{naturemag}
\bibliography{sci,arXiv,multifi}